\RequirePackage{fix-cm}
\documentclass[natbib,smallextended]{svjour3}       
\smartqed  
\usepackage{graphicx}
 \usepackage{amsmath, amssymb, txfonts}
 \journalname{Space Sci. Rev.}
\newcommand{\rsun}{R_{\odot}}
\begin{document}

\title{Magnetic Flux Transport at the Solar Surface}


\author{J. Jiang         \and
        D. H. Hathaway   \and
        R. H. Cameron    \and
        S. K. Solanki    \and
        L. Gizon         \and
        L. Upton
}


\institute{J. Jiang \at
              Key Laboratory of Solar Activity, National Astronomical Observatories,
              Chinese Academy of Sciences, Beijing 100012, China \\
              \email{jiejiang@nao.cas.cn}
           \and
           D. H. Hathaway \at
              NASA/MSFC, Huntsville, AL 35812, USA \\
              \email{david.hathaway@nasa.gov}
           \and
           R. H. Cameron \at
              Max-Planck-Institut f\"ur Sonnensystemforschung,
              Justus-von-Liebig-Weg 3, 37077 G\"ottingen, Germany \\
              \email{cameron@mps.mpg.de}
             \and
           S. K. Solanki \at
              Max-Planck-Institut f\"ur Sonnensystemforschung,
              Justus-von-Liebig-Weg 3, 37077 G\"ottingen, Germany \\
              and School of Space Research, Kyung Hee University,Yongin,
              Gyeonggi-Do,446-701, Korea\\
              \email{solanki@mps.mpg.de}
           \and
           L. Gizon \at
              Max-Planck-Institut f\"ur Sonnensystemforschung,
              Justus-von-Liebig-Weg 3, 37077 G\"ottingen, Germany \\
              and Georg-August-Universit\"at G\"ottingen, Institut f\"ur Astrophysik,
              Friedrich-Hund-Platz 1, 37077 G\"ottingen, Germany \\
              \email{gizon@mps.mpg.de}
           \and
           L. Upton \at
              Vanderbilt University, Nashville, TN 37235 USA \at
              The University of Alabama in Huntsville, Huntsville, Al 35899 USA \\
              \email{lar0009@uah.edu}
}

\date{Received: date / Accepted: date}

\maketitle

\begin{abstract}
After emerging to the solar surface, the Sun's magnetic field
displays a complex and intricate evolution. The evolution of the
surface field is important for several reasons. One is that the
surface field, and its dynamics, sets the boundary condition for the
coronal and heliospheric magnetic fields. Another is that the
surface evolution gives us insight into the dynamo process. In
particular, it plays an essential role in the Babcock-Leighton model
of the solar dynamo. Describing this evolution is the aim of the
surface flux transport model. The model starts from the emergence of
magnetic bipoles. Thereafter, the model is based on the induction
equation and the fact that after emergence the magnetic field is
observed to evolve as if it were purely radial. The induction
equation then describes how the surface flows -- differential
rotation, meridional circulation, granular, supergranular flows, and
active region inflows -- determine the evolution of the field (now
taken to be purely radial). In this paper, we review the modeling of
the various processes that determine the evolution of the surface
field. We restrict our attention to their role in the surface flux
transport model. We also discuss the success of the model and some
of the results that have been obtained using this model.

\end{abstract}

\section{Introduction}
\label{sec:intro}

The magnetic fields on the Sun are generated by dynamo action, ultimately driven
by convective motions beneath the Sun's surface \citep{Charbonneau10}. Many of the
physically important dynamo processes take place beneath the solar surface,
where the details are mostly hidden from us. The tools we have for probing the subsurface
dynamics of the magnetic fields are theory and helioseismology, both of which have
unveiled some of the dynamics \citep[for a review of helioseismic results see][]{Gizon05}.

Our knowledge of the magnetic field dynamics at the solar surface
can be inferred from high resolution spectropolarimetric
observations, for example, the {\em Hinode} spacecraft with about
230 km resolution \citep{Tsuneta08} and the {\em Sunrise}
balloon-borne solar observatory with about 100 km resolution
\citep{Solanki10}, and is consequently much richer in detail. The
magnetic field at the solar surface is observed to be structured on
all spatial scales we can observe - from below the resolution limit
of the largest available solar telescopes to the scale of the whole
Sun \citep{Solanki+06}. In this review, we will concentrate
exclusively on the evolution of the large-scale magnetic fields at
the solar surface.

One reason for studying the evolution of the large-scale magnetic
field on the solar surface is because that it sets the structure of
the heliospheric magnetic field \citep{Mackay12}. A second reason is
that it is the observable part of the solar dynamo. In the context
of the Babcock-Leighton dynamo \citep{Babcock61, Leighton64}, the
surface evolution is particularly important because the source of
poloidal flux in this model is the emergence and subsequent
evolution of tilted magnetic bipolar regions.

The evolution of the surface magnetic field is, in its simplest
form, almost trivial. Magnetic flux emerges at the solar surface in
the form of bipolar magnetic regions. The flux is then transported
and dispersed over the solar surface due to systematic and turbulent
motions. Lastly, when magnetic flux of opposite polarity come into
contact, the features cancel, removing equal amounts of flux of each
sign.

These processes are modeled by the surface flux transport equation,
which describes the evolution of the radial component of the
magnetic field $B_r$ on the solar surface. The equation is the
$r$-component of the MHD induction equation at $r=R_{\odot}$ under
the assumption that the field at the surface is purely vertical,
augmented by a source term for $B_r$, and flux removal term,  $S$ and
$D$ respectively \citep[see][]{DeVore_etal84} . The equation for the
radial component of the field  $B_r$ at  $r=R_{\odot}$ is then
\begin{eqnarray}
\frac{\partial B_r}{\partial t}=
     &-&\frac{1}{R_{\odot}\sin \theta} \frac{\partial } {\partial
     \phi}  \left(u B_r \right)
      -\frac{1}{R_{\odot}\sin \theta} \frac{\partial}{\partial \theta}
     \left({\varv}
           B_r \sin\theta \right) \nonumber \\ \noalign{\vskip 2mm}
     &+& \eta_H\left[\frac{1}{R_{\odot}^2\sin \theta}
     \frac{\partial}{\partial \theta}
     \left(\sin \theta \frac{\partial B_r}{\partial \theta} \right)
         + \frac{1}{R_{\odot}^2\sin^2 \theta}\frac{\partial^2
     B_r}{\partial \phi^2}\right] \nonumber \\ \noalign{\vskip 2mm}
    &+& D(B_r) + S(\theta,\phi,t),
\label{eqn:SFT}
\end{eqnarray}
where $u(\phi,\theta,t)$ is the velocity in the longitudinal
($\hat{\phi}$) direction, $\varv(\phi,\theta,t)$ is the velocity in
the latitudinal ($\hat{\theta}$) direction, $\eta_H$ is the
horizontal diffusivity at the surface (which we have assumed is
uniform), $D$ is some operator representing the removal of flux from
the surface, and $S$ is a source term describing the emergence of
new flux rising from below, $\phi$ and
$\theta$ are the solar longitude and colatitude respectively
and $R_{\odot}$ is the solar radius.

In principle, both the the surface velocity,
$u(\phi,\theta,t)\boldsymbol{\hat{\phi}}
 +\varv(\phi,\theta,t)\boldsymbol{\hat{\theta}}$,
and the radial component of the magnetic field are structured on all
scales from tens of meters to the size of Sun, and evolves on time
scales of seconds for the small scales to years for the largest
scales. This renders the full problem intractable. For almost all
problems, however, the full range of scales do not need to be dealt
with, and average values of $u$ and $\varv$ can be used, with
smaller unresolved velocities being treated as an enhanced
diffusivity $\eta_H$. There is no single best choice of what
temporal or spatial averaging should be done: different temporal and
spatial averaging allow different science questions to be addressed.

In the following sections we will add flesh to Eq. (1) by describing
in detail the relevant physical processes and the ways in which they
can be modeled. We start with a deeper exposition of the basis for
the surface flux transport model in Section 2. Then we describe some
of the ways in which the source term $S$ can be constructed in
Section 3, and the flows and diffusivity in Section 4. The removal
of the magnetic flux from the solar surface is reviewed in Section
5.  The results from using the surface flux
transport model will be presented in Section 6. Section 7 concludes
our review.

\section{Observational Basis for Solar Surface Flux Transport}
\label{sec:passiveTransport}

The part of the magnetic field at the Sun's surface that dominates
the signal in magnetograms, such as those recorded by the MDI
instrument \citep{Scherrer95} on SOHO or by the HMI instrument
\citep{Scherrer12, Schou12} on SDO, is thought to be produced by a
dynamo that resides deep in the solar convection zone or in the
convective overshoot layer below the convection zone
\citep[e.g.][]{Weiss+Thompson09, Charbonneau10}. The toroidal field
concentrated there becomes buoyantly unstable once it reaches a
critical strength and a part of it, thought to be in the form of
magnetic flux tubes, rises through the convection zone until it
reaches the solar surface \citep{Parker55, Choudhuri+Gilman87,
Schuessler+94}. On the way to the surface, the rising magnetic flux
tube is affected by solar rotation (via the Coriolis force) and
convection, which affect its path and hence the longitudes and
latitudes at which the field finally emerges. See \cite{Fan09} for a
review. The combined effects of solar rotation and convection are
also responsible for the orientation of two polarities at the solar
surface (e.g., Joy's law) \citep{Weber11,Weber13}.

With its footpoints simply thought to remain connected with the
horizontal toroidal magnetic field, the rising flux tube becomes
akin to an $\Omega$-shaped magnetic loop. The top of this loop is
the first feature to appear above the solar surface. Its footpoints
at the solar surface move apart rapidly as lower parts of the loop
reach the solar atmosphere.

By the time the magnetic flux tube reaches the surface, it has
typically been shredded into smaller features by the convection.
Hence, on small scales the emerging magnetic field initially
presents a  complex pattern on the solar surface \citep{Cheung+08}.
With time the many small magnetic structures partly grow together
again. This is particularly striking in the case of sunspots, which
often originally appear at the surface in the form of fragments that move
together, joining up to form the final, larger sunspot. Young
sunspots and active regions also display some amount of twisting
motion \citep{Brown+03}, which is thought to be associated with the
unwinding of the heavily twisted emerging magnetic loop.

Hence the horizontal motions associated with the early evolution of
the magnetic field after it reaches the solar surface mainly appear
to reflect its own internal dynamics, dictated by its rise and the
interaction of the flux tube with the convection in the solar
interior (as well as any unwinding that may happen in the process).
However, even while the emergence process is ongoing, other forces
start acting to move and shape the magnetic field at the solar
surface.

Once at the surface the magnetic field is  affected by a number of
large- as well as small-scale flows. These include differential
rotation \citep{Howe09} and its variation in the form of torsional
oscillations \citep{Howard80}, meridional circulation
\citep{Miesch05, RightmireUpton_etal12, Zhao_etal13}, and different
scales of convection ranging from granulation \citep{Nordlund09} to
supergranulation \citep{Rieutord+Rincon10} and possibly larger
scales \citep[e.g.,][]{Hathaway+13}. More about the large-scale and
small-scale flows will be given in Section \ref{sec:processes}.

That these flows can drag along the magnetic field is related to the high
magnetic Reynolds number, $R_m = U L/\eta$, where $U$ is a
typical flow velocity, $L$ the length scale of the flow and $\eta$
is the molecular magnetic diffusivity (which is inversely proportional to the
electrical conductivity). In and on the Sun, at the scales we are interested in,
we have $R_m \gg 1$, so that the magnetic field is frozen into the gas \citep{Choudhuri98}.

How strongly the horizontal components of the various flows at or
close to the solar surface move the magnetic elements depends on
both the strength of the flows relative to the strength of the
magnetic field and how strongly the features are anchored below the
surface. A critical quantity is the equipartition field strength,
$B_{\rm eq} = \sqrt{4\pi \rho}v$, where $\rho$ is the gas density
and $v$ is the magnitude of the velocity of the (convective) flow.
Magnetic fields that are weak compared to $B_{\rm eq}$ will always
be basically dragged by the flows, whereas stronger fields can
influence the flows if they are anchored below (which requires $B
\ge B_{\rm eq}$ all the way down to their anchoring depth). The
expectation is then that the magnetic elements will move with some
(weighted) average of the velocity field over the range from where
it is anchored. It has been argued that even large, strong-field
features at the solar surface, such as sunspots, lose the connection
with their roots at the bottom of the convection zone at rather
shallow depths \citep{schuessler+rempel05}. The simulations by
\citet{Rempel11} indicated that the anchoring depth, which ranges
from few Mm to dozens of Mm, is related to the lifetime of the
sunspot.

On the Sun we have the interesting situation that while averaged
over the solar disk the field strength is well below the
equipartition value, the individual strong-field magnetic features
have kG fields \citep[e.g.,][]{Solanki+06}. This makes their fields
considerably stronger than $B_{\rm eq}$, which is around 200-400 G
\citep{Solanki96} in the lower photosphere for granular flows and
smaller for slower flows (such as of supergranulation). The
strong-field magnetic features, i.e., magnetic elements, pores and
sunspots, make up the dominant part of the field seen in most
magnetograms.

It turns out that the size of the magnetic features helps determine
whether they affect the flow or are moved by it. Thus, sunspots are
located at the centers of moat cells and pores also have a positive
divergence of horizontal velocity surrounding them
\citep{Verma+Denker14}. Smaller magnetic features, however, are
almost always situated at the edges of convection cells. In the
quiet Sun the magnetic field forms a network at the edges of
supergranules, while in active regions the structuring is generally
on a mesogranular scale \citep{DominguezCerdena03}. On a smaller
scale magnetic elements are found almost exclusively at the edges of
granules \citep{Title87, Solanki89}. Hence observationally it is
clear that the magnetic field is dragged along by convective flows
on different scales. The effect of the meridional circulation is
difficult to determine well from direct measurements (see Section
\ref{MC}) due to the slow speeds of a few ms$^{-1}$ (but plays an
important role in flux transport computations; see Section
\ref{sec:model}). The fact that the strong-field (i.e., kG) magnetic
features are mostly aligned radially \citep[i.e., vertically in the
local solar coordinates,][]{Martinez_Pillet97, Jafarzadeh14b}, makes
it easier for the field to be advected passively.

Studies of the motion of individual magnetic features show that
these resemble a random-walk  process, with the features moving
between granules as these grow, evolve, move and die. On a larger
scale these motions are affected by the location of the magnetic
features within the supergranules, being subdiffusive in regions of
converging supergranular flows and superdiffusive in the bodies of
supergranules \citep{Abramenko_etal11,  Jafarzadeh14}.

Strong evidence that magnetic features are advected along with
horizontal flows on the solar surface comes from the comparison of
results from surface flux transport simulations  with the observed
distributions of magnetic fields. More about surface flux transport
models will be given in the upcoming sections of this paper.

\section{Sources of Magnetic Flux}
\label{sec:sources} In this section, we begin our description of the
individual physical processes relevant to the evolution of the
large-scale magnetic field on the Sun's surface. We begin with
flux-emergence which is the process that brings magnetic field
generated by dynamo action through the solar surface. The largest
scales of emergence are large active regions with length scales on
the order of 100~Mm and fluxes of $\sim 6\times 10^{22}$~Mx. They
are observed to extend down to the smallest scale loops currently
observable \citep{Centeno_etal07, Ishikawa_etal10} with fluxes of
$10^{17}$~Mx, based on {\em Hinode} observations, and the almost
ubiquitous emergence found by \cite{HagenaarCheung09} and
\citet{Danilovic10} using {\em Hinode} and {\em Sunrise}
observations respectively. Below currently resolvable limits,
recirculation of magnetic fields and dynamo action in the turbulent
intergranular lanes are believed to occur \citep[][and references
therein]{deWijn+09}.

The emergence processes have been modeled in detail for both
large-scale active regions \citep[e.g.,][]{Cheung+08, Cheung_etal10,
Stein_etal11} and for the small-scale dynamo processes
\citep{Voegler+Schuessler07, SchusslerVogler08}. The physics
involved includes magnetic buoyancy, magnetic tension, gravity,
radiative cooling, thermodynamics including the effect of partial
ionization, and small-scale turbulence which drains mass from the
loops \citep[see e.g.,][]{Cheung+08}.

This review does not deal explicitly with intranetwork fields (the
weak field that lies inside the superganular network), nor with the
even smaller scale, more turbulent field found in the quiet Sun by
the Hanle effect. See \cite{deWijn+09} for a review of quiet-Sun
fields. The evolution of such a field at the solar surface is
expected to be different from that of the field produced by a global
dynamo, given that the intranetwork field is relatively weak and
horizontal \citep{Lites08,Jin09}, and hence is transported even more
easily by convective flows. It is easily deformed and distributed by
the turbulent convection, so that distinct magnetic features lose
their identity relatively quickly.

The model to understand the solar surface flux transport process does not include the physics
necessary to properly describe the evolution of the field during emergence, which are intrinsically
three dimensional. Rather, the model assumes that the emergence occurs on a time scale much shorter
than those otherwise of interest, enabling the emergence to be treated as occurring instantaneously.
The source term for one particular emergence event (event $i$) therefore
has the form $S_i(\theta,\phi,t) = S_i(\theta,\phi) \delta(t-t_i)$.
The prescription of $S_i$ is not unique in the literature, and depends on the purpose
of the study and the observational data that are available to reconstruct $S_i$.
Ordering them by the extent to which they include the details of observations of
individual emergence events, the different ways of creating $S_i$ are
\begin{enumerate}
\item{Replacing magnetic fields at low latitudes by observations \citep[e.g.,][]{DurrantMcCloughan04}.
This is a type of data assimilation. }
\item{Magnetogram based sources \citep[e.g.,][]{Yeates_etal07}.}
\item{Sunspot areas and locations, together with an empirically derived law to convert the areas to
fluxes, with  Joy's law \citep{Sheeley_etal85},
or a cycle-dependent version of Joy's law \citep{Cameron_etal10}, or the observed tilt angles of the
individual groups.}
\item{Sunspot numbers, with the properties of the sunspots group based on random realizations of
empirically derived distributions  \citep[e.g.,][]{Schrijver_etal02,
Jiang_etal11b}.}
\item{Empirical laws \citep[e.g.,][]{vanBallegooijen_etal98}.}
\end{enumerate}

For those methods that do not simply rely on magnetic field
assimilation (i.e., methods 2--5 in the above list),
$S_i(\theta,\phi)$ represents an isolated bipolar magnetic region,
usually the superposition of positive and negative polarity patches
displaced some distance from one another. The most important
physical constraint on $S_i$ is that the total (signed) flux
vanishes over some small distance. This requirement follows from the
induction equation
\begin{equation}
\frac{\partial {\bf{B}}}{\partial t}=\nabla \times \left( {\bf{U}} \times {\bf{B}} \right)
       -\nabla \times \left( \eta \nabla \times {\bf{B}} \right)
\end{equation}
applied to a local patch of the solar surface $\Sigma$. By Stokes'
theorem we have

\begin{equation}
\int_{\Sigma} \frac{\partial {\bf{B}}}{\partial t} \cdot
\mathbf{\hat{n}}~\mathrm{d}{\Sigma}=
       \int_{\partial \Sigma} \left( {\bf{U}} \times {\bf{B}}
                           -\eta \nabla \times {\bf{B}} \right) \cdot  \mathrm{d}{\bf{l}},
\end{equation}

where $\partial \Sigma$ is the boundary of $\Sigma$ and
$\bf{\hat{n}}$ is the unit vector normal to the surface element
$\mathrm{d}{\Sigma}$. This reduces to
\begin{equation}
\frac{\partial}{\partial t} {\int_{\Sigma} B_r \mathrm{d}{{\Sigma}}} =\int_{\partial \Sigma}
             \left( {\bf{U}} \times {\bf{B}}
                         -\eta \nabla \times {\bf{B}} \right) \cdot  \mathrm{d}{\bf{l}},
\end{equation}
from which it can be seen that the only way the magnetic flux
integrated over any region of the solar surface $\Sigma$ can change
is by advection or diffusion across the boundary of the region
$\partial \Sigma$ \citep[the argument given here is similar to that
in][]{Durrant_etal01}. For truly instantaneous emergence, the
opposite polarities must balance over a very small region. For
emergence taking place over a day, the flux must be balanced on
scales of about $\sim 100$~Mm (this being the distance field can be
carried by a 1~km s$^{-1}$ flow over the course of a day).

Usually, each bipolar magnetic region is idealized as a pair of equal and opposite fluxes
concentrated around the centroid of their respective polarities. Also, each such doublet
is typically emerged suddenly at the time that its flux is largest. The contribution of
the magnetic flux to the surface field is
\begin{equation}
S_i(\theta,\phi)=B^{+}_i(\theta,\phi)-B^{-}_i(\theta,\phi),
\end{equation}
where $B_i^{\pm}$ is the flux distribution of the positive and
negative polarity of the $i$-th bipolar magnetic region (BMR). Two
major methods have been developed to give these distributions. One
is from the NRL group, e.g, \cite{Sheeley_etal85}, \cite{DeVore87}
and \cite{Wang_etal89} who took each region as a point bipole. It
has the form
\begin{equation}
B_i^{\pm}(\theta,\phi)=\frac{\Phi_i\delta(\theta-\theta_i^{\pm})\delta(\phi-\phi_i^{\pm})}{R^2_\odot\sin\theta_i^{\pm}},
\end{equation}
where $\Phi_i$ is total flux of the BMR and $(\theta^\pm,\phi^\pm)$
are the co-latitude and longitude of each polarity of the BMR. The
other method was initiated by \cite{vanBallegooijen_etal98} and was
adopted by others \citep{Mackay_etal02a, Mackay_etal02b,
Baumann_etal04, SchusslerBaumann06, Cameron_etal10, Jiang_etal11b,
UptonHathaway14}. Instead of point sources, they used finite-sized
Gaussian-like polarity patches. The areas, locations (latitude and
longitude), and latitudinal separations determined by the tilt
angles of BMRs determine the source flux distribution. Specific
details for sources used in many models and how the source
parameters affect the flux transport are given in
Section~\ref{sec:model}.

The long-term sunspot record from the network of observatories by
Royal Greenwich Observatory (RGO), starting in May of 1874 and until
1976 and continued by the Solar Optical Observing Network (SOON)
since 1976, provides daily observations of the location and area of
sunspot groups. The systematic differences in the area measurements
between the two datasets pose a barrier to understanding and
reconstructing the long-term magnetic field evolution. A factor of
about 1.4 was suggested to correct the SOON area to be homogeneous
with RGO data \citep{Balmaceda09}. Another disadvantage of RGO/SOON
data is the absence of information concerning the tilt angles. The
records of sunspots based on white-light photographs from the
observatories at Mount Wilson (MWO) in the interval 1917-1985
\citep{Howard_etal84} and at Kodaikanal in the interval 1906-1987
\citep{Sivaraman93} provide two large, but not complete, samples of
sunspot group tilt angles. These records are being extended based on
data from the Debrecen observatory \citep{Gyori11}. Magnetic
polarities of the sunspot groups cannot be identified from the
white-light photographs. The studies based on the magnetograms show
that sunspot groups have reversed polarity orientations (anti-Hale
source) with percentages ranging from 4\% to 10\% \citep{Wang89,
tian03, Stenflo12, Li12}.

\begin{figure*}
\begin{center}
\includegraphics[scale=0.5]{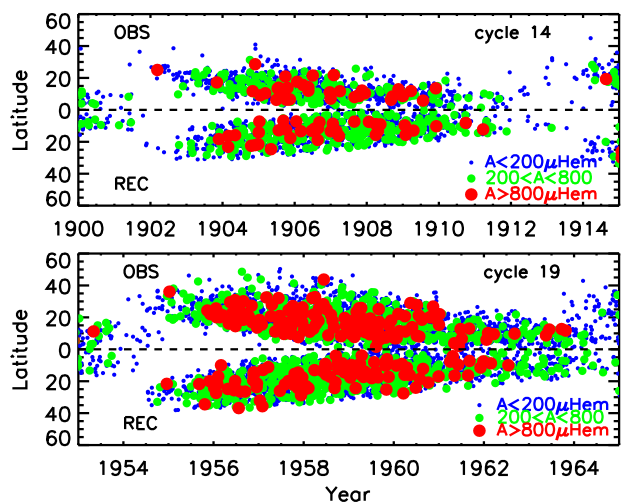}
\end{center}
\vspace{3mm} \caption{Comparison of butterfly diagrams from
observation (above the horizontal dashed lines) and reconstruction
(below the dashed lines) for the weakest cycle 14 covered by RGO
period (upper panel) and the strongest cycle 19 (lower panel), both
for the northern hemisphere. The area of the sunspot groups is
indicated by the colors and sizes of circles
\citep[from][]{Jiang_etal11a}.} \label{fig:ButterflyReconJiang11a}
\end{figure*}

The dependence of the statistical properties of sunspot emergence on
the cycle phase and strength may be derived using the historic
record of sunspot groups together with the group
\citep[$R_G,$][]{Hoyt98} or Wolf \citep[$R_Z$,][]{Wolf61} sunspot
number. Using the group sunspot number $R_G$ and RGO, MWO and
Kodaikanal data sets, the main correlations found are as follows.
(i) Strong cycles have a higher mean latitude for sunspot emergence
\citep{Waldmeier55,Solanki08}. The mean latitude at which sunspots
emerge can be modeled using a second order polynomial of cycle phase
\citep{Jiang_etal11a}. (ii) The distribution of sunspot areas is
similar for all cycles \citep{Bogdan88}. (iii) The size distribution
is a power-law for small sunspots \citep{Baumann05} and obeys a
log-normal profile for large sunspots \citep{Bogdan88}. During cycle
maxima, sunspots are larger on average \citep{Jiang_etal11a}. (iv)
The cycle averaged tilt angle is anti-correlated with the cycle
strength \citep{DasiEspuig_etal10, Dasi-Espuig13}. (v) Sunspot nests
are important, especially during cycle maximum phases. Using these
empirical characteristics, the time-latitude diagram of sunspot
group emergence (butterfly diagram) was reconstructed by
\citet{Jiang_etal11a} from 1700 onward on the basis of the Wolf and
group sunspot numbers. Figure \ref{fig:ButterflyReconJiang11a} shows
the comparison of butterfly diagrams from observation and
reconstruction for the weakest cycle 14 covered RGO period (upper
panel) and the strongest cycle 19 (lower panel), both for the
northern hemisphere.

\section{Flux Transport Processes}
\label{sec:processes}

For any particular scale at the surface, the flows that transport
the magnetic flux can conveniently be categorized as systematic
flows or random motions. This distinction is only possible once the
spatial and temporal scales relevant to the study have been decided.
At scales below those that we are interested in, random flows with
zero mean can be treated in several ways, as discussed below. The
systematic flows include the differential rotation and the
meridional circulation.

The random-walk effect introduced by the random flows can be treated
as diffusion with a diffusivity estimated from the observed motions
of the magnetic elements or the characteristics of the convective
flows themselves. The differential rotation and meridional
circulation can both be measured using a variety of techniques,
including feature tracking, direct Doppler measurements, and
helioseismology. A wide range of studies have been carried out to
investigate the natures of the flux transport processes, which are
reviewed in the following subsections.

\subsection{Diffusion}
\label{sec:diffusion}

One of the key terms of the flux transport is the horizontal diffusion of the radial component of the field.
The Spitzer value for the magnetic diffusivity in the solar photosphere becomes relevant on scales of 30~km
for a time scale of one day, which is a much smaller scale than the surface flux transport (SFT) model aims to capture.
On the scales of interest, which are much larger than 30~km, there is a choice as to how to treat the
random flows.

One approach, adopted by \cite{Schrijver01} and
\cite{UptonHathaway14}, is to include in the advection velocities
and small-scale cellular flows or random motions corresponding to,
e.g., supergranulation. The second, more commonly used approach, is
to model the small-scale random motions as a turbulent diffusivity,
$\eta_H$. The value of $\eta_H$ is therefore not the Spitzer
diffusivity, but rather a parameterization of the effect of the
turbulent near-surface convective motions on the magnetic field.

The initial estimation of $\eta_H$ by \cite{Leighton64}, based on the
correct reversal time of the polar fields without including
meridional flow, was in the range 770\,--\,1540 km$^2$s$^{-1}$. The
value was lowered to around 200\,--\,600 km$^2$s$^{-1}$ once
meridional flow was included \citep{DeVore_etal84}. \cite{Mosher77}
derived a value of 200\,--\,400 km$^2$s$^{-1}$ using magnetic
observations to trace the history of a typical solar active regions.
Using similar methods, \cite{SchrijverMartin90} estimated a
diffusivity of about 250 km$^2$s$^{-1}$ in a quiet region
surrounding a magnetic plage, and 110 km$^2$s$^{-1}$ in the magnetic
plage itself. The results from a number of observational studies are
summarized in Table 1 of \cite{Schrijver_etal96}. Values of $\eta_H$
between 100 and 340~km$^2$ s$^{-1}$ have been found on spatial
scales in the 6 Mm range using comprehensive photospheric
simulations with different upper boundary conditions
\citep{Cameron_etal11}, and values of $\sim 100$~km$^2$ s$^{-1}$
based on a mean-field motivated analysis of numerical simulations
and {\em Hinode} data \citep{Rudiger_etal12}. The photospheric
motions responsible for the turbulent diffusion range from
turbulence in the intergranular lanes, through granular motions to
supergranulation. Each of these types of motion occupies a range of
spatial scales, and $\eta_H$ in principle should therefore be a
function of spatial scale $k$ \citep{Chae_etal08, Abramenko_etal11,
Abramenko13}, with the issue being complicated by the limited
lifetime of the features being tracked and realization noise
\citep{Jafarzadeh14}. The values used in simulations cover the range
suggested by observations, and a parameter study of the effects of
varying $\eta_H$ was reported by \cite{Baumann_etal04} and is
discussed further in Section~\ref{sec:model}.

\subsection{Differential Rotation}
\label{sec:DR}

The Sun's differential rotation is the oldest known, and best
characterized, flux transport process. It has a dynamic range of 250
m s$^{-1}$ in latitude and a well characterized latitudinal and
radial structure thanks to  helioseismology. The near-surface radial
shear is also of importance for the magnetic flux transport as the
magnetic elements are anchored within this layer. See also
\citet{Beck00} for a review.

\begin{figure}[ht!]
\includegraphics[width=\columnwidth]{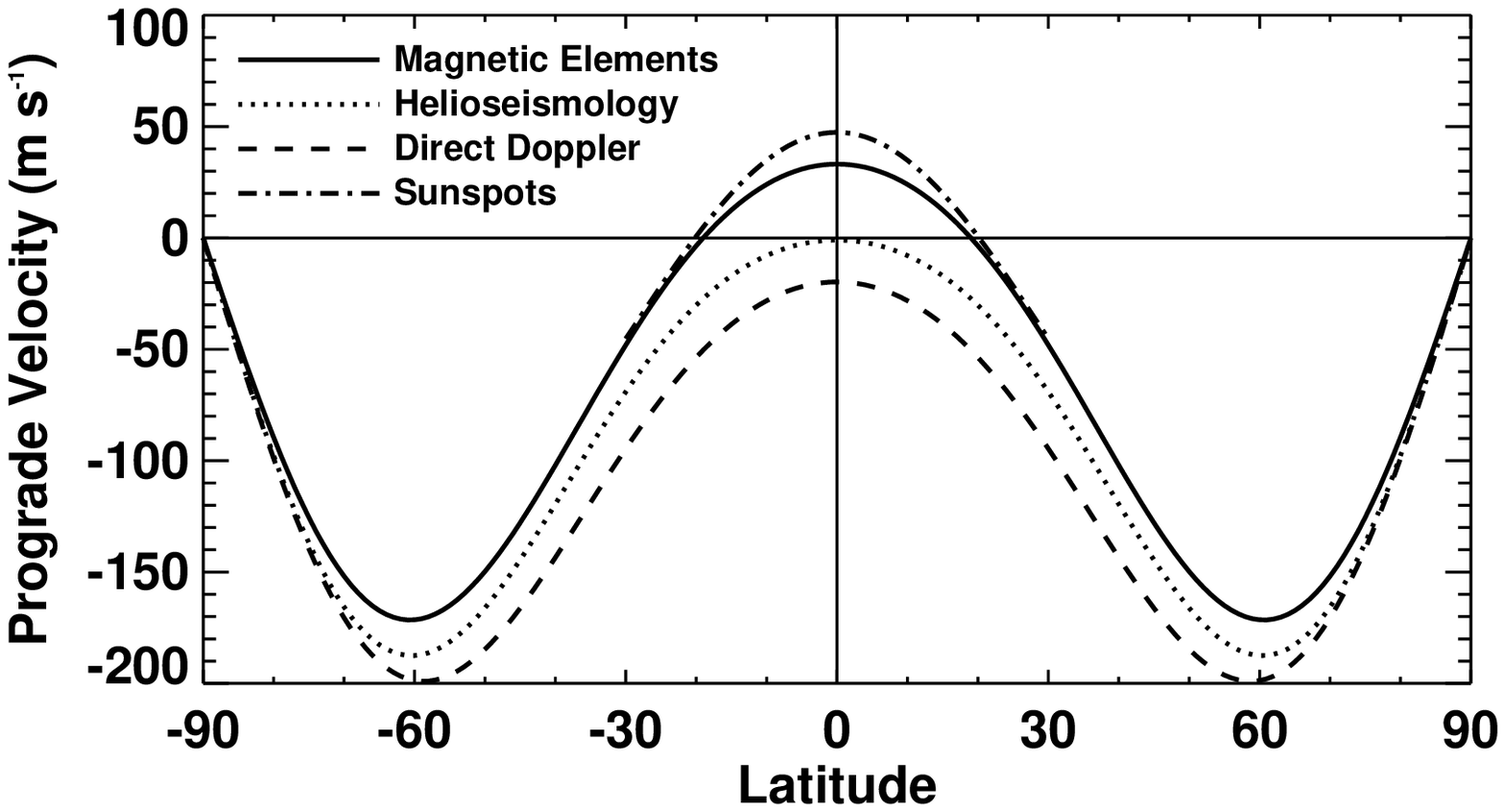}
\caption{Differential rotation profiles as measured by different
methods. The profile from the small magnetic elements measured by
\citet{Komm93a} is given by the solid line. The profile from
global helioseismology at $r=0.995 \rsun$ measured by
\citet{Schou_etal98} is given by the dotted line. The profile from
direct Doppler measured by \citet{Ulrich_etal88} is given by the
dashed line. The profile from individual sunspots measured by
\citet{Howard_etal84} is given by the dashed-dotted line. The zero line represents solid body Carrington rotation.}
\label{fig:DRprofile}
\end{figure}

The motions of sunspots gave the first measure of the latitudinal
differential rotation (first noted by Christoph Scheiner in 1610),
with well-characterized rotation profiles given by
\citet{NewtonNunn51}, by \citet{Ward66}, and by
\citet{Howard_etal84}. These rotation profiles only cover the low
latitudes (30$^\circ$ and below) and they indicate that spots of
different sizes have different rotation rates (faster rotation for
smaller spots). The rotation profile derived for all spots by
\citet{Howard_etal84} is indicated by the dashed-dotted line in
Fig.~\ref{fig:DRprofile}.

Direct Doppler measurements \citep{HowardHarvey70, Snodgrass_etal84, Ulrich_etal88} extend to all latitudes.
These measurements indicate a slower rotation rate in the photosphere.
The average profile measured by \citet{Ulrich_etal88} is plotted in Fig.~\ref{fig:DRprofile} as a dashed line.

Global helioseismology \citep{Thompson_etal96, Schou_etal98} gives a surface shear layer in which, at low to moderate latitudes,
the rotation rate increases inward from the photosphere to a depth of about 50 Mm or 7\% of the solar radius.
This shear layer is clearly seen in the lower latitudes but its structure becomes more uncertain at latitudes
greater than about 50$^\circ$ \citep{CorbardThompson02}.
Local Helioseismology gives similar results \citep{Giles98,Basu_etal99, Komm_etal03} that also indicate uncertainty at
the higher latitudes. The profile obtained with global helioseismology by \citet{Schou_etal98} at $r = 0.995 \rsun$
(a depth of 3.5 Mm) is plotted with the dotted line in Fig.~\ref{fig:DRprofile}.

The motions of the small magnetic elements \citep{Komm93a,
Meunier05, HathawayRightmire10, Hathaway11} show a similar shape of
the differential rotation profile, but substantially faster rotation
speeds than those given by direct Doppler measurements in the
photosphere or from helioseismology at a depth of 3.5 Mm. The
profile obtained by \citet{Komm93a} is plotted with the solid line
in Fig.~\ref{fig:DRprofile} and is given by

\begin{equation}
u(\theta) = (33 - 281 \cos^2 \theta - 293 \cos^4 \theta) \sin \theta
\mathrm{~~} {\mathrm{ms}}^{-1}
\end{equation}

\noindent where $u(\theta)$ is relative to the Carrington frame of reference.

\begin{figure}[ht!]
\includegraphics[width=\columnwidth]{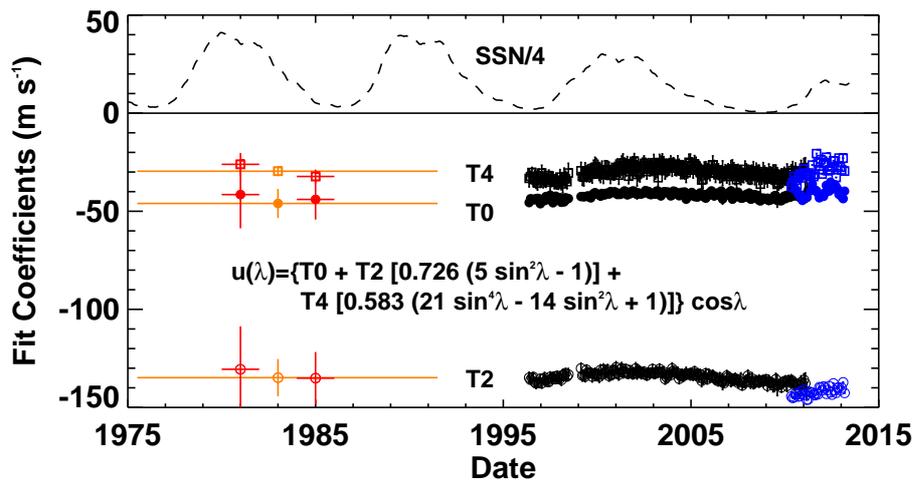}
\caption{The recent history of the polynomial fit coefficients for
differential rotation profiles as measured by the motions of the
magnetic elements. The coefficient T0 (giving solid body rotation)
is represented by filled circles. T2 is represented by open circles
and T4 by open squares. The \citet{Komm93a} measurements for
1975-1991 are shown in orange. Cycle 21 maximum (1980-1982) and
cycle 21/22 minimum (1984-1986) are shown in red. The
\citet{Hathaway11} measurements for individual Carrington
rotations (1996-2010) are shown in black while recent results
obtained from SDO/HMI measurements are shown in blue.}
\label{fig:DRprofileHistory}
\end{figure}

The surface differential rotation varies over the course of each sunspot cycle in small but systematic ways.
Changes in the overall shape of the differential rotation can be followed by tracking the changes in the coefficients that fit the profiles.
Care should be taken, however, to cast the fits to the profiles in terms of orthogonal polynomials (in this case associated Legendre polynomials of order 1) as was suggested by \citet{Snodgrass84} to avoid crosstalk between the coefficients.
Results of doing this for the measurements made with the small magnetic features are shown in Fig.~\ref{fig:DRprofileHistory}.
The average values obtained by \citet{Komm93a}, for the length of their study (1975 to 1991), are shown in orange with 1$\sigma$ error bars for the first three north-south symmetric polynomials
(given by the expression included within the figure).
\citet{Komm93a} also provided coefficients for cycle 21 maximum (1980-1982) and for cycle 21/22 minimum (1984-1985).
These are shown in red with 1$\sigma$ error bars.
The results for individual Carrington rotations, obtained from SOHO/MDI magnetograms by \citet{Hathaway11}, are shown in black with 2$\sigma$ error bars.
This is augmented by results from SDO/HMI magnetograms shown in blue with 2$\sigma$ error bars.

\begin{figure}[ht!]
\includegraphics[width=\columnwidth]{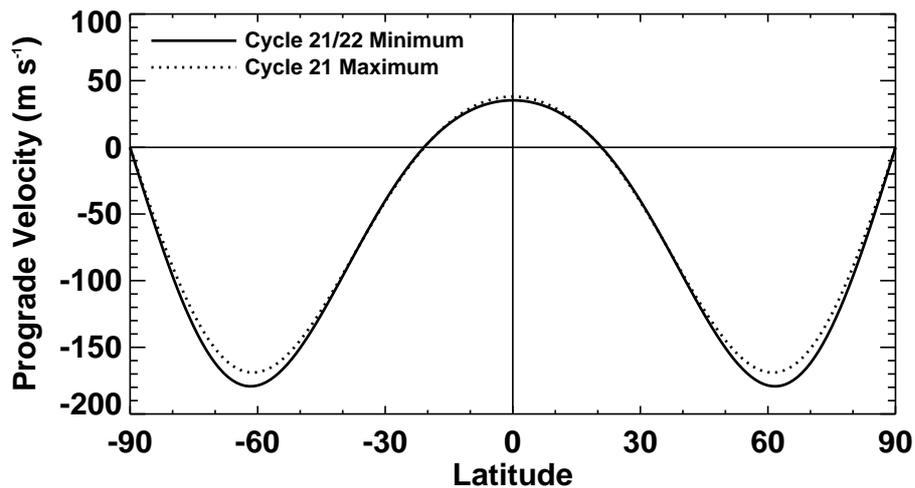}
\caption{Differential rotation profiles at sunspot cycle minimum and
maximum. The profile for cycle 21/22 minimum (1984-1986) from
\citet{Komm93a} is represented by the solid line. The profile for
cycle 21 maximum (1980-1982) by the dotted line. }
\label{fig:DRminMax}
\end{figure}

All three coefficients are smaller (in absolute terms) at sunspot cycle maxima than they are at cycle minima.
This gives a slightly faster (less negative relative to the Carrington rate) solid body rotation but a weaker differential rotation with less latitudinal shear at cycle maxima.
The differences in the differential rotation flow profiles between cycle minima and maxima are nonetheless quite small as shown in Fig.~\ref{fig:DRminMax}.

In addition to these systematic changes to the basic profile there
are the smaller scale, evolving perturbations referred to as
torsional oscillations by \citet{Howard80}. These variations in the
differential rotation profile are easily seen after removing an
average profile \citep{Howe11}. The deviations from the average
profile are in the form of latitude bands with faster and slower
than average rotation rates. The faster bands are located on the
equatorward sides of the sunspot zones, while the slower bands are
located on the poleward sides. This system of fast and slow streams
drifts equatorward with the sunspot zones but are apparent at higher
latitudes years before sunspots appear. While these flows are
clearly associated with the solar cycle and are of considerable
interest, the relative flows are quite weak ($\sim 5$ m s$^{-1}$)
and thus probably of little consequence for surface flux transport.
The torsional oscillations are also seen with helioseismology
\citep{Schou_etal98} and extend in depth throughout the convection
zone \citep{Vorontsov02}. In addition, helioseismology revealed the
existence of a second torsional oscillation branch, which propagates
poleward, at high latitudes \citep{Schou99}.

\subsection{Meridional Circulation}
\label{MC}

A meridional flow was implicated in surface flux transport long
before its strength (or even direction) had been well-determined. In
his pioneering paper on the solar dynamo \citet{Babcock61} suggested
that there was a meridional circulation that spread outward from the
active latitudes. In his model, the higher latitude poleward flows
would transport following polarity flux to the poles where it would
reverse the polar fields halfway through the cycle and then build up
new polar fields with the sign of the following polarity in each
hemisphere. Babcock's model also included a low latitude equatorward
flow that would transport preceding polarity flux to the equator,
where it would cancel with the opposite polarity from the other
hemisphere. This meridional flow seemed reasonable based on the
effects of the Coriolis force on the differential rotation relative
to the Carrington rotation frame of reference -- the low latitude
faster flow would be turned equatorward by the Coriolis force while
the high latitude slower flow would be turned poleward. Babcock also
cited observations of the motions of sunspots which suggested a
meridional flow of this form. After the ``discovery'' of
supergranules by \citet{Leighton_etal62}, Babcock's meridional
circulation was deemed unnecessary by \citet{Leighton64} who
proposed that the surface flux transport was all done by a random
walk of the magnetic elements due to evolving granules and
supergranules.

The earliest measurements of the meridional flow were based on the
motions of sunspot groups. \citet{DysonMaunder13} used sunspot group
motions to refine the determination of the orientation of the Sun's
rotation axis and noted a tendency for high latitude groups to move
poleward and low latitude groups to move equatorward.
\citet{Tuominen42} examined the meridional motions of recurring
sunspot groups (groups that live long enough to be identified on
more than one disk passage) and found that these groups did indeed
diverge from the active latitudes with velocities of 1-2 m s$^{-1}$.
Similar results were found for individual sunspots by
\citet{HowardGilman86}. There are two significant problems in using
the meridional motions of sunspots in surface flux transport models:
sunspots do not appear at high latitudes (thereby leaving the
meridional flow unknown poleward of about 40$^\circ$) and the motion
of sunspots may not representative of the surface meridional flow.

\begin{figure}[ht!]
\includegraphics[width=\columnwidth]{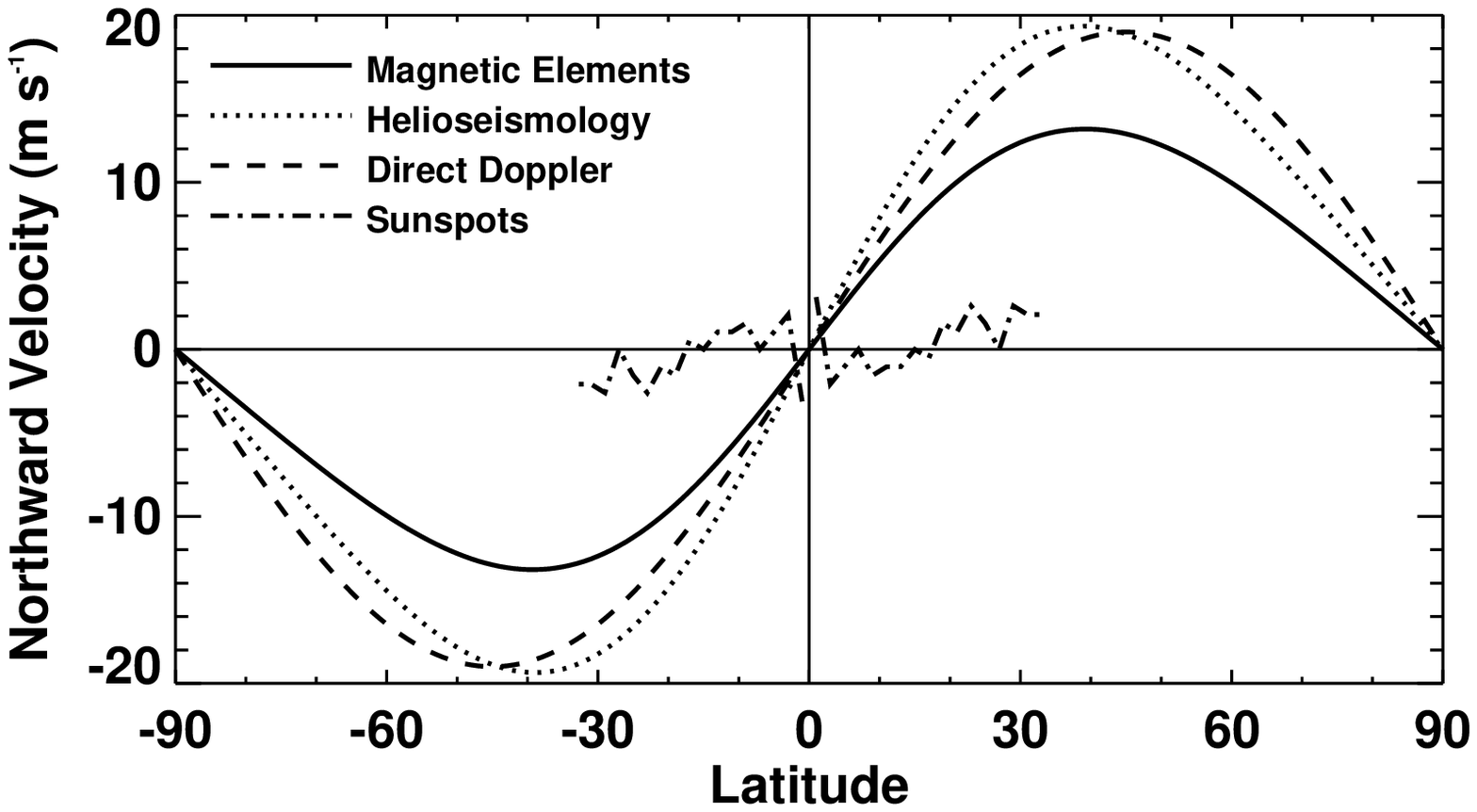}
\caption{Meridional flow profiles as measured by different methods.
The profile from the small magnetic elements measured by
\citet{Komm93} is given by the solid line. The profile from
local helioseismology at $r=0.998 \rsun$ measured by
\citet{BasuAntia10} is given by the dotted line. The profile from
direct Doppler measured by \citet{Hathaway96} is given by the dashed
line. The profile from recurrent sunspot groups measured by
\citet{TuominenKyrolainen82} is given by the dashed-dotted line.}
\label{fig:MFprofile}
\end{figure}

More or less complete latitude coverage is available with direct Doppler, helioseismology, and feature
tracking using the small magnetic elements that populate the entire surface of the Sun.
Figure~\ref{fig:MFprofile} shows some of the meridional flow profiles that have been reported.
There are small but significant differences in the surface velocity derived from the different techniques.
This is partly because the measurements are all subject to systematic uncertainties
and sample different depths.

\begin{figure*}
\begin{center}
\includegraphics[scale=0.32]{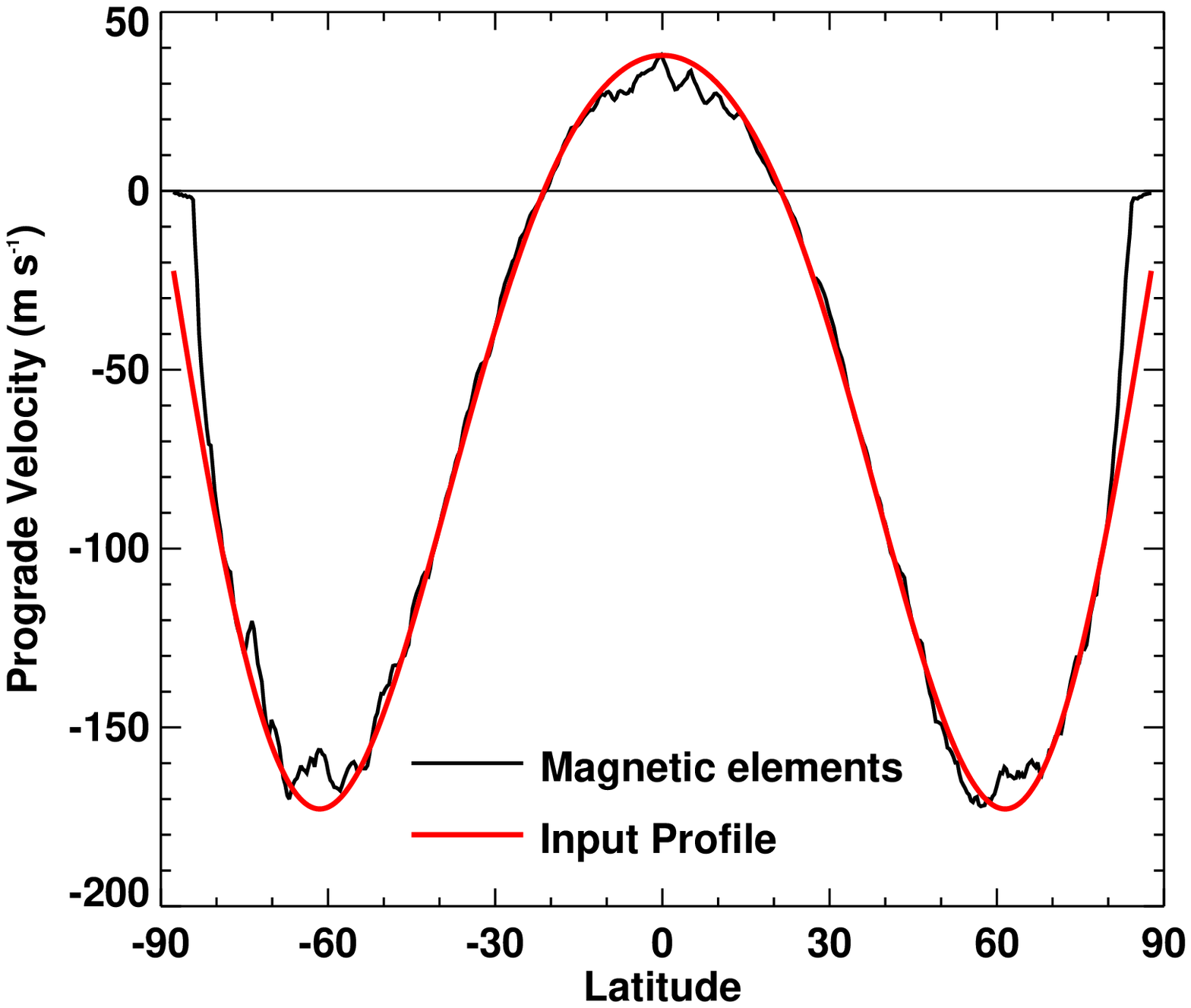}
\includegraphics[scale=0.32]{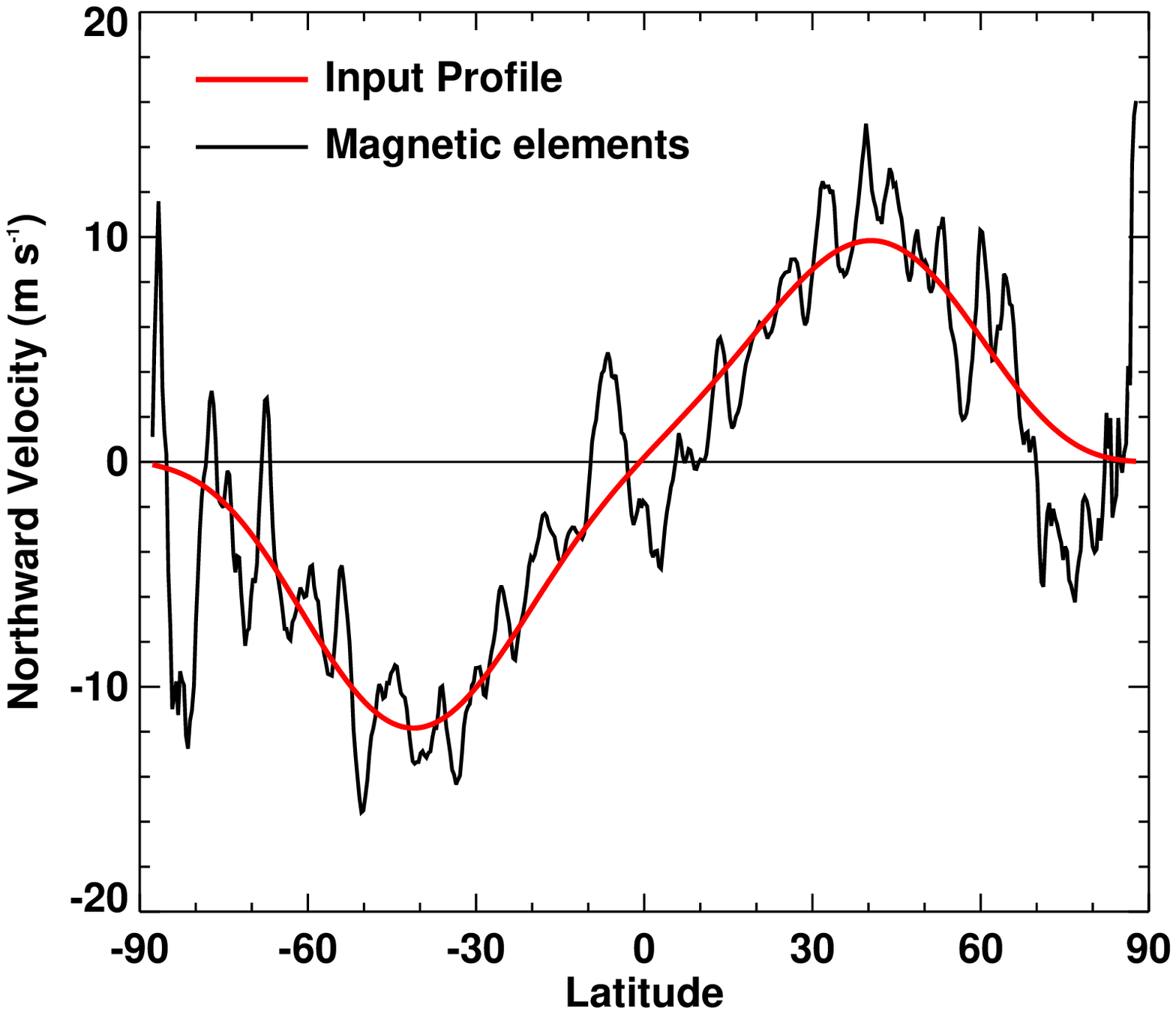}
\end{center}
\caption{Profiles of the differential rotation (left) and meridional
flow (right) for the maximum of cycle 23 (calendar year 2000). The
profiles input to the fully advective flux transport model of
\citet{UptonHathaway14} are shown in red. The profiles measured
using magnetic element feature tracking on this data are shown in
black.}
\label{fig:FlowValidation}
\end{figure*}

Measurements of the near-surface meridional motions of the small magnetic elements
\citep{Komm93, Gizon03, HathawayRightmire10, Hathaway11, RightmireUpton_etal12}
can also be used to determine the meridional velocity.
A typical cycle-averaged meridional flow profile determined by magnetic feature tracking is as given by
 \cite{Komm93} as
\begin{equation}
v(\theta) = (31.4 \cos \theta - 11.2 \cos^3 \theta) \sin \theta
\mathrm{~~} {\mathrm{ms}}^{-1}.
\end{equation}
\citet{Hathaway11} tested the sensitivity of magnetic feature
tracking as a way of determining the large-scale flows to the
effects of the random motions of the  magnetic elements. They took a
magnetic map representative of cycle maximum, represented the
magnetic field distribution on a 4096-by-1500 grid in longitude and
latitude by a collection of some 120,000 magnetic elements that were
then advected by an evolving pattern of supergranules. They did not
find any substantial flow away from the active latitudes as was
suggested by \citet{Dikpati10}. They later \citep{UptonHathaway14}
produced a fully advective surface flux transport code in which the
magnetic elements are transported by the flows in an evolving
pattern of supergranules. They assimilate data from magnetograms on
the Sun's near side but the field evolution on the far side is
produced purely by the surface flux transport. Figure
~\ref{fig:FlowValidation} shows that both the differential rotation
and meridional flow measured by magnetic element feature tracking on
the far side data for the maximum of cycle 23 (the year 2000) do not
differ significantly from the input profiles for this choice of
random (supergranular) motions. They argue that the velocity field
determined in this way is the most consistent for use in the SFT
model.

The interpretation of the Doppler measurements are complicated by
the presence of the strong convective blue shift signal
\citep{Hathaway96, Ulrich10}. This signal is an apparent blue shift
in spectral lines due to the correlation between emergent intensity
and radial flows in granules. It can vary by as much as 500 m
s$^{-1}$ between disk center and limb, and is affected by the
presence of magnetic field (\citet{Welsch13} studied the Doppler
velocity details in active regions and noted that the presence of
magnetic fields can have substantial effects on the observed Doppler
velocities.). The Doppler signal from the meridional flow has a
spatial structure similar to that of the convective blue shift but
with a maximum of only 10 m s$^{-1}$ -- hence the difficulty in
measuring the meridional flow from the Doppler shift of spectral
lines.

\begin{figure*}
\begin{center}
\includegraphics[scale=0.6]{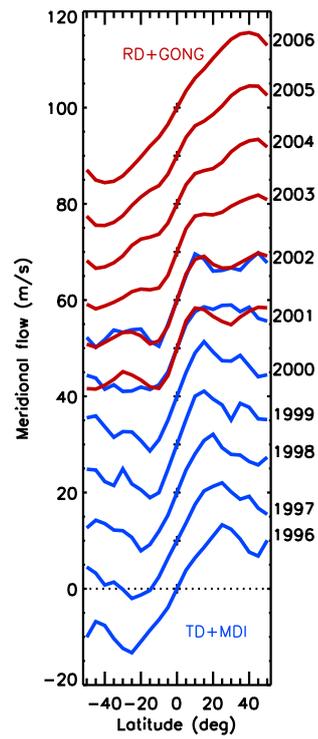}
\end{center}
\vspace{0.3cm} \caption{Meridional circulation measured by local
helioseismology in the near-surface layers and its evolution from
1996 to 2006. Individual years are shifted by multiples of
10~ms$^{-1}$ for clarity. Blue curves show the results from
\cite{Gizon08} (time-distance measurements of the advection of the
supergranulation pattern using SOHO/MDI data). Red curves show the
results from \cite{Gonzalez08} (ring-diagram measurements using GONG
data and multiplied by a factor of 0.8 to match the blue curves in
the years 2001 and 2002). Only the antisymmetric components with
respect to the equator are shown. This figure is taken from
\cite{Gizon10}.} \label{fig:ARAA}
\end{figure*}

Measurements of the meridional flow have been made using several
local helioseismic techniques, with similar results for the
near-surface flows. The first such measurement \citep{Giles_etal97}
used the method of time-distance helioseismology
\citep{Duvall_etal93} and gave a poleward flow of approximately
$\sim 20$ m s$^{-1}$ at 30$^{\circ}$. More recent near-surface
measurements, covering most of cycle 23, are shown in Fig.
~\ref{fig:ARAA} from two different techniques: MDI time-distance
helioseismology measurements of the advection of the
supergranulation pattern \citep{Gizon03,Gizon08} and GONG
ring-diagram helioseismology \citep{Gonzalez08}. The peak
near-surface meridional velocity is about 15~ms$^{-1}$ at a latitude
of $\sim$30$^{\circ}$ in these more recent studies.

The time dependence of the meridional flow is clearly seen in the
local helioseismology observations. We see in Fig.~\ref{fig:ARAA}
that, in the rising phase (1996 to 2002) of the cycle, the latitude
where the meridional circulation peaks moves towards lower
latitudes. The time-varying component of the near-surface meridional
flow is consistent with an inflow into the active latitudes
\citep{Gizon04,Zhao04,Gizon08,Gonzalez10}. The inflows into
individual active regions can be seen in two dimensional maps, first
reported by \cite{Gizon01} using $f$-mode time-distance
helioseismology. A theory for the inflows, related to the enhanced
cooling associated with the bright plage, was suggested by
\cite{Spruit03} with a demonstration of the plausibility of the
idea being given in \cite{Gizon08}.


Using magnetic feature tracking applied to MDI observations, \cite{Meunier99}
detected clear changes in the meridional flow associated with active regions.
The more extensive MDI measurements of \cite{HathawayRightmire10} show changes
with the solar cycle indicated by: 1) polynomial fits to the profiles (Fig. 8) and 2)
by detailed changes to the meridional flow profiles \citep{Hathaway11}
fully consistent with superimposed inflows toward the active regions
\citep{Cameron10}.

\section{Sinks of Magnetic Flux}
\label{sec:sinks}

Without the supply of new flux introduced by $S$, the total unsigned
flux at the solar surface monotonically decreases. For plausible
values of the meridional flow speed and magnetic diffusivity, the
e-folding time of the slowest decaying solution is about 4000 years
\citep{Cameron07}. The slowest decaying solution consists of equal
amounts of oppositely directed magnetic flux that is well
separated, concentrated at the two poles. Much more rapid decay
occurs when the two polarities are close to each other, with an
e-folding time of at most a few years as the field of both
polarities is advected to the poles where the two polarities then
come into close proximity and cancel. The cancellation is, in most
models, due to the diffusion term $\eta_H \nabla^2 B_r$, i.e. it is
due to magnetic reconnection which is assumed to occur in the
photosphere.

The second type of sink is represented by the term $D(B_r)$. This
type of term was introduced by \cite{Schrijver_etal02} and
\cite{Baumann_etal06}. In physical terms, the idea is that processes
below the surface of the Sun, where magnetic diffusion is also
operating, cause the magnetic field at the surface to decay in-situ
\citep{Baumann_etal06}. Because both sources and sinks are subject
to the same requirement that changes in flux must be localized,
\cite{Schrijver_etal02} suggests the possibility that the decay of
the field is accompanied by the emergence of a large number of
small, weak bipoles that together form a chain of loops, and allows
the field to appear to decay in-situ without violating the argument
presented in Section~\ref{sec:sources}. As opposed to emergence
events, the decay envisaged here is slow, and the fluxes are low, so
that the observational signal of the large chains of bipoles can be
lost in the noise of the omnipresent flux recycling.

Because $D(B_r)$ is a parametrisation of the physics of the highly
dynamic convection zone, both its functional form and amplitude are
open to discussion. \citet{Baumann_etal06}, for example, consider
that the functional form of $D(B_r)$ might reflect the eigen
solutions of the problem of free-decay in a static convection zone
with a uniform diffusivity, which fixes the functional form for the
diffusivity. \cite{Schrijver_etal02}, on the other hand, consider a
simpler model where the field decays with a constant $e$-folding
time.

Once the functional form of $D(B_r)$ is chosen, the question of its
amplitude arises. Again this is, in principle, difficult to
determine from first principles as it depends on the properties of
the turbulence in  the convection zone, with mean-field
magnetohydrodynamic effects such as turbulent diffusivity and
magnetic pumping playing a role. The strength of $D(B_r)$, in those
cases where it has been included, is chosen to ensure that the polar
fields reverse during each cycle.

In practical terms, $D(B_r)$ reduces the 4000 year memory of the SFT
model to a few cycles. Such a reduction of the memory of the system
might be physically justified \citep[as suggested
by][]{Schrijver_etal02, Baumann_etal06}, but also has the effect of
removing the long term accumulation of small errors in the modeling.
For example, \citet{Jiang_etal11b} used $D(B_r)$ to reduce the
effects of the imperfect knowledge and therefore modeling of the
source term $S$.

We comment that nonlinearities can be included in the model via the
cycle dependence of the latitude and tilt angle at which sunspot
groups emerge \citep{Cameron_etal10,Jiang_etal11b}, or the global
meridional circulation rate \citep{Wang_etal02a}, or localized
inflows into the active region latitudes \citep{CameronSchussler12}.
Depending on the time being simulated, these nonlinearities can
remove the need for a diffusive term to ensure the cyclic reversal
of the polar fields and a match with observations.

\section{Solar Surface Flux Transport Models}
\label{sec:model}
\subsection{A reference model}
\label{sec:Standard model} In the above sections, we have presented
the observational features of the solar surface flow and the surface
flux source due to the BMR emergences. \cite{BabcockBabcock55}
speculated that the following flux of BMRs tended to migrate
poleward, while the leading flux tended to migrate equatorward. The
poleward migration of following flux would neutralize and reverse
the solar polar fields over the course of a sunspot cycle.
\cite{Babcock61} later speculated that the observed poleward
migration might reflect a pattern of meridional flow on the Sun.
\cite{Leighton64} proposed an alternative mechanism -- that the
random motions of magnetic flux by supergranular flows, together
with Joy's law would lead to a preferred equatorward diffusion of
leading flux and poleward diffusion of following flux. This proposal
did not require other latitudinal transport mechanisms. However,
\citet{Mosher77} showed that the diffusivity needed by Leighton was
much higher than suggested by the observed flows and that a
systematic flow was required. From the 1980s onwards such
large-scale flows, including differential rotation and meridional
flows have been included in the models
\citep{DeVore_etal84,Sheeley_etal85,DeVore87,Wang_etal89}. A
historical review of the development of the surface flux transport
model has been given by \citet{Sheeley05}. The models and
applications of the magnetic flux transport at the solar surface
flux were also reviewed by \citet{Mackay12}.

The SFT model (described by Eq.~(\ref{eqn:SFT})) has been applied to
the evolution of the Sun's global field \citep[see][ and numerous
papers there after]{DeVore_etal84}. It has also been applied
\citep[e.g.,][]{Schrijver01} to smaller scales, from large active
regions to small ephemeral regions. It has been applied by treating
the supergranular motions as a diffusivity, as well as by explicitly
modeling them \citep{UptonHathaway14}.

For the differential rotation, the synodic rotation rate of the large-scale magnetic field,
as measured by \cite{Snodgrass83}, is widely used in SFT models. It is
\begin{equation}
\label{eq:st_rotation}
\Omega(\theta)=13.38-2.30\cos^2\theta-1.62\cos^4\theta-13.2
\textrm{~~deg~day$^{-1}$}.
\end{equation}

For the meridional flow, the profiles
\begin{equation}
\label{eq:st_mf} \upsilon(\theta)=31.3 \mid
\sin\theta\mid^{2.5}\cos\theta  \mathrm{~~} {\mathrm{ms}}^{-1}
\end{equation}
and
\begin{equation}
\label{eq:st_mf2} \upsilon(\theta)=\left\{
  \begin{array}{l l}
     11\sin\left[2.4\ast(90^{\circ}-\theta)\right] ~\mathrm{ms^{-1}} & \mathrm{where}~ 15^{\circ}< \theta < 165^{\circ}\\
     0 & \textrm{otherwise},
  \end{array}
  \right.
\end{equation} are close to the solid curves in Fig.\ref{fig:DRprofileHistory} at middle
and low latitudes. In contrast, a sharp gradient near the equator was used by
\citet{Wang_etal89,Wang_etal09}. The comparisons of the different profiles are shown in Fig. 11 of
\cite{Hathaway11} and Fig. 3 of \cite{Jiang_etal13}.

As a reference model, we take the transport equation
Eq.~(\ref{eqn:SFT}), with the source in the form of used by
\cite{vanBallegooijen_etal98}, transport parameters, i.e.,
meridional flow and differential rotation in the forms of
Eq.~(\ref{eq:st_rotation}) and (\ref{eq:st_mf}), 250 km$^2$s$^{-1}$
horizontal turbulent diffusivity, and zero radial diffusivity.

\subsection{Evolution of an Individual Sunspot Group:
Effects of Different Model Parameters} \label{sec:BMR_simulations}

The axisymmetric component of the large-scale field is measured by the axial dipole
moment, which is defined as
\begin{equation}
D_{\mathrm{Axial}}(t)=\frac{3}{4\pi}\int B(\theta,\phi,t)\cos\theta\sin\theta d\theta
d\phi.
\end{equation}
\label{eq:dipole}
In this review, we do not consider the equatorial dipole
field, which is strongly affected by differential rotation and hence
has a short life time, on order of one year \citep{DeVore87}.
We note that although such a field is not central for the solar
cycle evolution of the Sun's surface field,
it is an important ingredient in the evolution of
solar open flux \citep{Mackay_etal02a,WangSheeley02}.

\subsubsection{Source Parameters}
\label{sec:source_parameters} The initial contribution of an
individual BMR with tilt angle $\alpha$ and total flux $F$ (area
$A$) located at colatitude $\theta$, to the solar axial dipole field
may be expressed as
\begin{equation}
\label{eq:dipole_bmr} D_{\mathrm{BMR}}\propto d
F\sin\theta\sin\alpha,
\end{equation}
where $d$ is the distance between the opposite polarities. The axial
dipole of the bipole then evolves due to the latitudinal transport
of the two polarities, which depends on  both diffusion and flows.
In the presence of diffusion alone, the axial dipole field decays on
a time scale $\tau_d/2=\frac{1}{2}R_\odot^2/\eta_H$
\citep{Leighton64, Baumann_etal06}, which is approximately 30~years
for a diffusivity of 250 km$^2$s$^{-1}$. For the pure advection
case, the dipole field is proportional to $\sin\theta$ and
declines on the time scale $\tau_\upsilon \sim R_\odot/\upsilon_0
\sim 11$~years \citep{WangSheeley91} as both polarities are swept to
the poles. In the presence of both systematic flows and diffusion
(or random motions) a fraction of the magnetic field can cross the
equator (under the action of the diffusive or random motions) after
which they are kept apart by the meridional circulation.

Left panel of Fig. \ref{fig:dip_BMR_lati} from \citet{Jiang14} shows
the combined effect of diffusion and flow on the axial dipole field
of a single BMR with area 1000 $\mu$Hem, total flux
6$\times10^{21}$Mx and a large tilt angle of 80$^\circ$ emerged at
different latitudes (latitudes 40$^\circ$, 30$^\circ$, 20$^\circ$,
10$^\circ$ and 0$^\circ$). The BMRs at the high latitudes and close
to the equator display quite different dipole field evolution. For
the cross-equator emergence (0$^\circ$), the centroids of the two
polarities are located at about $\pm4.3^\circ$. Advection in each
hemisphere separates the polarities and causes the increase of the
dipole field. Part (about half) of the flux diffuses and annihilates
across the equator along the polarity inversion line
\citep{Mackay_etal02a}. The remaining flux eventually concentrates
around the poles and the dipole field reaches a plateau.
\citet{Jiang14} show that the equilibrium axial dipole field
generated by the emergence of a single such extreme cross-equatorial
BMR is about 20\% of the total simulated dipole field generated by
all recorded sunspots groups of cycle 17, which had a medium
amplitude. When the BMR emerges at 10$^\circ$ and 20$^\circ$, the
poleward flow gradient (larger gradient at lower latitudes) causes
an increase of the separation between the polarities and an increase
of the dipole field during the beginning phase. Then more leading
flux is transported to the same pole and annihilated with the
following polarity. This causes a weaker equilibrium field for a BMR
emerging at higher latitude. For BMRs emerging at 30$^\circ$ and
40$^\circ$, the dipole field diminishes in about 2 years. The right
panel shows the relation between the final axial dipole field and
the latitudinal location of the BMR with a given magnetic flux and
tilt angle. The solid curve represents a Gaussian fit with a HWHM in
latitude of 8.8$^\circ$.

Hence, the large BMRs with large tilt angles emerging close to the
equator contribute most to the solar axial dipole field. Usually the
BMRs are assumed to obey the Hale's polarity law in the SFT models.
The anti-Hale spots generate the same amplitude of the axial dipole
field as the spots obeying Hale's law, but have opposite sign.

\subsubsection{Transport Parameters}
\label{sec:transport_parameters} Differential rotation is one of the
key ingredients in the evolution of the non axisymmetric component
of the large-scale magnetic field \citep{DeVore87}. It has no
effects on the axial dipole field. Hence we do not discuss its
effects here.

Figure \ref{fig:dip_para} shows the dependence of the BMR's axial
dipole fields  after reaching equilibrium on the diffusivity (left
panel) and on the maximum meridional flow strength (right panel).
The BMR has 1000 $\mu$Hem area, 6$\times10^{21}$Mx total flux and
normal tilt angle 5$^\circ$. We deposit the BMR at 8$^\circ$ (dashed
line) and 18$^\circ$ (solid line) to show the different effects of
the diffusion and meridional flow on the BMR eruptions at different
latitudes. The reference model is used except for the variations of
the diffusivity and the meridional flow strength.

The BMR located at high latitude (18$^\circ$) generates
higher equilibrium dipole fields at higher diffusivity since more
flux from the leading polarity can diffuse across the equator and
be transported into the opposite hemisphere. For the BMR
located at low latitude (8$^\circ$), the axial dipole field
increases with the increase of the diffusivity when the diffusivity
is low. When the diffusivity is further increased, more flux will be
canceled between the two opposite polarities, which causes the
decrease of the equilibrium dipole fields. The BMR at a latitude of 8$^\circ$
always generates a stronger dipole field than that at
18$^\circ$ latitude.

The axial dipole field monotonically decreases to zero with increasing meridional
flow when the BMR is deposited at 18$^\circ$
latitude. This is because more leading polarity flux is transported to the
same pole as the following polarity due to the stronger
meridional flow. When the flow is strong enough, all the leading
polarity flux is transported to the north pole without diffusion
across the equator. When the BMR is deposited at 8$^\circ$ latitude,
being close to the equator facilitates cross-equator diffusion.
When the flow strength is low, more of the flux cancels
before the equilibrium dipole field is established. Increasing flow speed
decreases the flux cancellation and hence generates a stronger axial
dipole field. When the flow is further increased, the flux diffusing
across the equator decreases. Hence the axial dipole field
decreases. This numerical simulation implies that the variation of the
meridional flow might have different effects on the axial dipole field evolution of
different cycles since the latitudinal distribution of the sunspot groups depends on
the cycle strength \citep{Solanki08,Jiang_etal11a}.

The effects of perturbations to the meridional flow in the  form of
inflows toward the active latitudes, as described in
Section~\ref{MC}, on the evolution of solar surface axial dipole
field was studied by \cite{Jiang_etal10}. In each hemisphere, an
axisymmetric band of latitudinal flows converging toward the central
latitude of the activity belt was superposed onto the background
poleward meridional flow. The overall effect of these flow
perturbations is to reduce the latitudinal separation of the
magnetic polarities of a BMR and thus diminish its contribution to
the equilibrium axial dipole field.

\subsection{Simulations of Solar Cycles}
\label{sec:cycle_simulations}

\subsubsection{Comparisons of Observed and Simulated Magnetic Butterfly Diagrams}

In his original paper on the transport of solar magnetic flux,
\cite{Leighton64} simulated the effect of the thousands of sources
that occur during an entire sunspot cycle. Cycle 21 was the first
cycle that permitted a realistic comparison with the observed field
\citep{Sheeley_etal85,DeVoreSheeley87,Wang_etal89}. The observed
features of BMRs were derived from the full-disk magnetograms. The
large-scale axisymmetric magnetic field features, such as the polar
field structure, poleward surges and polar field reversals were well
reproduced. The time evolution of the longitudinally averaged
photospheric magnetic field, i.e, the magnetic butterfly diagram, is
a good illustration of the large-scale field evolution under the
flux transport process.

The upper panel of Fig. \ref{fig:MagbutterflyObsSim} shows the
magnetic butterfly diagram resulting from a flux transport simulation, the
source and transport parameters of which are based on \citet{Jiang10a}, see
also \citet{SchusslerBaumann06}. The lower panel of Fig.~\ref{fig:MagbutterflyObsSim} is
produced from the Kitt Peak Solar Observatory
synoptic magnetograms of the radial magnetic field. There are
qualitative agreements between simulation and observation,
particularly concerning the poleward surges of following-polarity
magnetic flux leading to the reversals of the polar fields.

Some differences can also be identified between the simulated and
the observed magnetic butterfly diagrams. For example, the
observations have a more grainy structure, which leads to a high
mean flux density at the activity belt, see Eq.(9) of
\citet{Jiang14} for the definition. The average of the observed
values over the three cycle maxima is about 3G, which is about twice
that of the simulated result. Furthermore, the simulations lack the
occasional cross-equatorial flux plumes that appear in the data due
to the large, highly tilted  sunspot groups that emerge near the
equator, for example in the years of 1980, 1986, and 2002
\citep{Cameron_etal13}.

The differences can mainly be attributed to the scatter in sunspot
group tilt angles relative to Joy's law. \citet{Jiang14} measured
the tilt scatter based on the observed tilt angle data from MWO and
Kodaikanal. The standard deviations ($\sigma_\alpha$) of the tilt
angles depend on the sunspot area in the form of $\sigma_\alpha=-11
\log_{10}(A_u)+35$, where $A_u$ is the umbra area.
Figure~\ref{fig:MagbutterflyCy17OnOffTiltScatter} shows the
comparisons of the simulated magnetic butterfly diagrams using the
observed sunspot records of cycle 17, which is a cycle with an
average strength and not associated with a sudden increase or
decrease with respect to the adjacent cycles, without (upper panel)
and with (lower panel) the tilt scatter.  The lower panel
corresponds to one random realization of the sunspot group tilt
scatter, which generates the similar polar field as the upper panel
without the tilt scatter. The randomly occurring large tilt angles
cause the more grainy structure, which is represented by an increase
of the low latitude flux density by about 40\% compared to the case
without tilt angle scatter. There are also more poleward surges with
opposite polarities. Qualitatively, the magnetic butterfly diagram
for the cases with tilt angle scatter is more similar to the
observed counterpart for the last 3 cycles. See \citet{Jiang14} for
more details about the effects of the scatter in sunspot group tilt
angles on the magnetic butterfly diagram. Occasionally, the near
equator sunspot groups with big sizes have big tilt angles.
According to Section \ref{sec:source_parameters}, a single such
event can significantly affect the axial dipole field at the end of
the cycle. If the event obeys the Hale polarity law, it strengthens
the axial dipole field. If the event is anti-Hale, it weakens the
axial dipole field.

\subsubsection{Simulations of Multiple Solar Cycles}

The success of the SFT model, with BMR emergence as the main source
of flux, opens the possibility for the reconstruction of the solar
large-scale magnetic field into the past on the basis of recorded
sunspot data. The observed cycle-to-cycle variations provide
constraints for the modeling of the different physical processes in
the model. When the BMR source amplitude fluctuations were included
in the model, \citet{Wang_etal02a} and \citet{Schrijver_etal02}
found that the polar field cannot reverse polarity every $\sim$11
yr. In their studies, the BMRs of different cycles had the same
range of latitude distributions. The tilt angles of BMRs obeyed
Joy's law and did not depend on the cycle strength. The total
intrinsic axial dipole field was proportional to the total flux of
the emergent sunspot groups during a cycle. Under the same transport
parameters, the strength of the polar field then varied linearly
with the total amount of emerged flux. During the weaker cycles the
flux supply was insufficient to cancel the existing polar field, to
reverse it and to build up a new polar field of opposite polarity
and of the same strength as before. Three different ways of
resolving this discrepancy have been put forward.

$\bullet$ \emph{Including in $D$ a component due to the
intrinsically three-dimensional nature of flux transport}

The reference SFT model described in Section \ref{sec:Standard
model} is explicitly two dimensional. With $S=D=0$, there is no flux
transport across the solar surface. In models simpler than the SFT,
multi-year decay times were proposed by \cite{Solanki00} to
successfully describe the evolution of the total amount of open and
total magnetic flux. \citet{Schrijver_etal02} and
\citet{Baumann_etal06} introduced different forms of $D(B_r)$ in
order to account for an intrinsically three-dimensional decay of the
field. \citet{Schrijver_etal02} found that a simple exponential
decay of the field $\tau_d$ with a decay time of about 10 yr allowed
regular reversals of the polar fields given fluctuations in the
source term similar to those in the historical records.
\citet{Baumann_etal06} introduced a more detailed expression for
$D(B_r)$ based on a parameterization of radial diffusion processes.
A radial diffusivity of 100 km$^2$s$^{-1}$ (corresponding to a decay
time of $\sim$5 years for the dipole component) was suggested. See
Section \ref{sec:sinks} for more discussions.

$\bullet$ \emph{Nonlinearities in the transport parameters}

Variations in the meridional flow have been considered as an
alternate way of ensuring the polar fields reverse at the end of
each cycle. The two types of changes considered are a modulation of
the global flow speed \citep{Wang_etal02a, Wang_etal05}, or the
inclusion of a localized inflow into active regions
\citep{CameronSchussler12}. The model of the inflow in the latter
study was calibrated to helioseismic observations
\citep{Cameron10}, although more work is needed to assimilate the
raw observations into their model. Both types of nonlinearities can
lead to reversals of the polar fields at the end of each cycle.

$\bullet$ \emph{Nonlinearities in the source parameters}

In Section~\ref{sec:sources} we have listed the characteristics of
sunspot group emergence. Strong cycles have a higher mean latitude
(related to the Waldmeier effect; \citeauthor{Waldmeier55}
\citeyear{Waldmeier55}) and a lower tilt angle for sunspot emergence
\citep{DasiEspuig_etal10}. According to the results discussed in
Section \ref{sec:source_parameters}, both the latitudes and the
tilts of the source term can significantly modulate the polar field
generation. \cite{Cameron_etal10} made the first attempt to
introduce nonlinearities in the source parameters to study the
magnetic field evolution of multiple cycles. Figure
\ref{fig:PolarFieldFluxCJSS10} shows the average of the unsigned
polar field strength from the flux transport model (red) and
observed sunspot area (black). In agreement with observations, the
polar field at the end of a solar cycle is correlated with the
subsequent cycle strength \cite[e.g., see][]{Munoz-Jaramillo13}, and
similarly for the open flux \cite[e.g., see][]{Wang09}.

Mixed approaches are also possible. \citet{Wang03} include the
nonlinearities in both the source and the transport parameters to
simulate the evolution of the Sun's large-scale magnetic field under
Maunder minimum conditions. They showed that the regular polarity
oscillations of the axial dipole and polar fields can be maintained
if the source flux emerges at low latitudes ($\sim10^{\circ}$) and
the speed of the poleward surface flow was reduced from $\sim20$ to
$\sim10$ m s$^{-1}$. \cite{Jiang_etal11b} have used semi-synthetic
records of emerging sunspot groups based on sunspot number data as
input for a surface flux transport model to reconstruct the
evolution of the large-scale solar magnetic field from the year 1700
onward. A nonlinear modulation of the tilt angles and emergence
latitudes based on observations was included as well as a decay term
$D$ based on the formalism in \cite{Baumann_etal06} with
$\eta_r=25$~km$^2$s$^{-1}$ to reduce the error in the modeling due
to the errors in the sunspot numbers.
Figure~\ref{fig:PolarFieldJiang11} shows the reconstructed polar
field based on Wolf sunspot number during 1700-2010 from
\cite{Jiang_etal11b}.

\subsubsection{Assimilations of Observed Magnetograms}

Surface flux transport models have also been used to construct
synchronic magnetic maps (maps of the magnetic field over the entire
surface of the Sun for a given moment in time) for use in coronal
field extrapolations and space weather predictions. In the above
sections, the flux sources were idealized as magnetic dipoles
produced  by the emergence of BMRs. For synchronic map production,
observed magnetograms are assimilated into a SFT model that then
includes the magnetic field evolution on the far side of the Sun.
\citet{WordenHarvey00} used their flux transport model and the Kitt
Peak synoptic magnetograms to update unobserved or poorly observed
regions. \citet{SchrijverDeRosa03} assimilated SOHO/MDI magnetograms
within 60$^\circ$ from disk center into a SFT model with an duration
of 5.5 yr and temporal resolution of 6~hours. With this they were
able to approximate the evolution of the photospheric magnetic field
on the unobservable hemisphere, and thus obtain a continuously
evolving model of the surface field over the whole solar surface.
\cite{SchrijverLiu08} extended the study throughout the whole of
cycle 23 to further understand the large-scale transport of the
magnetic flux in the solar photosphere. \citet{UptonHathaway14}
assimilated magnetograms from both MDI and HMI to produce a
``baseline'' set of synchronic maps from 1996 to 2013 at a 15-minute
cadence for comparison with maps made with BMR sources. They found
excellent agreement and showed that predictions of polar field
reversals and the polar field strength at cycle minimum could be
made years in advance. \citet{McCloughanDurrant02} and
\citet{DurrantMcCloughan04} noted that flux transport produces and
requires synchronic maps rather than traditional synoptic maps and
care must therefore be taken when estimating transport parameters
from synoptic maps.

\citet{Yeates_etal07} used synoptic magnetogram data as the initial
condition and assimilated the emergence of new active regions into
the model throughout the course of the simulation to maintain the
accuracy of the simulated photospheric magnetic field over many
months. The simulations were coupled with simulations of the 3
dimensional coronal magnetic field to explain the hemispheric
pattern of the axial magnetic field direction in solar filaments
\citep{Yeates09}.

\subsection{Peculiar Cycle 23 Minimum}

The polar field at the end of cycle 23 was unexpectedly weak, which
caused the unusual properties of the polar corona, the open flux,
and the solar wind at that time, see \citet{Wang_etal09} and
\citet{Jiang_etal13} for more details. As shown in Fig.
\ref{fig:cy17vs23}, cycle 23 has a similar amplitude and shape as
cycle 17. However, the amplitudes of their subsequent cycles, cycles
24 and 18, are very different. The cycle strength is proportional to
the polar field at the end of the preceding cycle \citep{Jiang07,
Munoz-Jaramillo13}, which implies that cycles with similar
amplitudes can generate rather different amounts of polar flux at
the end of the cycles. This situation poses an interesting challenge
to surface flux transport models.

\cite{SchrijverLiu08}, \citet{Wang_etal09} and \cite{Jiang_etal13} simulated the
evolution of the photospheric field of cycle 23 using flux transport
models. Sunspot number data were used to determine the number of
BMRs emergence at a given time. These studies  could produce the
observed weak polar field strength by increasing the meridional flow
relative to the reference case.

\cite{Yeates14} simulated cycle 23 by inserting individual BMR with
properties matching those in observed Kitt Peak synoptic
magnetograms. They also found that their standard flux transport
model is insufficient to simultaneously reproduce the observed polar
fields and butterfly diagram during cycle 23, and that additional
effects must be added. The variations they considered include an
increase of the meridional flow to 35 ms$^{-1}$, decrease of the
supergranular diffusivity to 200 km$^2$s$^{-1}$, decrease of the
sunspot groups tilt angle by 20\%, decrease of the flux per sunspot
groups by 20\%, inclusion of $D$ in Eq. (1) with a decay time of 5
years, decrease of the tilt angle of the sunspot groups by 20\%
coupled with radial diffusion in about 10 years, and the inflow
toward the active regions.

Stochastic variations in sunspot group emergence is another possible
cause of the weak cycle 23 minimum. As shown in Section
\ref{sec:source_parameters}, large highly tilted BMRs that emerge at
low latitudes produce cross-equatorial flux plumes in the synoptic
magnetograms and provide a large contribution to the axial dipole
field. \citet{Cameron14} simulated cycles 21-23 and showed that the
magnetic flux from four observed cross-equatorial flux plumes could
provide one explanation for the weakness of the polar fields at the
end of solar cycle 23.

\section{Conclusions}
\label{conclusions}

The solar photosphere is a thin layer between the high
plasma-$\beta$ solar interior and the low plasma-$\beta$ solar
atmosphere. It is the layer where the energy transport changes from
convective to radiative, the layer where the poloidal field is
generated in the Babcock-Leighton model and critically it is the
layer that we can observe and best measure the magnetic field. The
dynamics of the magnetic field in this layer are, based on
observations, particularly simple: emergence, dispersion and
advection by surface velocities, and eventually cancellation with
opposite polarity flux. These few processes explain the evolution of
the large-scale magnetic field at the solar surface, and beyond it
in the corona and the heliosphere. In this paper we have reviewed
these processes and shown how they can impact the evolution of the
Sun's magnetic field and the sunspot cycle.

The surface flux transport is the key to understanding what produces
the polar fields and the axial dipole moment seen at activity
minima. The strength of the polar fields at this phase of the
activity cycle is well correlated with the strength of the next
solar cycle and can be used as a reliable predictor
\citep{Schatten78, Schatten87, Svalgaard05, Jiang07, Wang09,
Munoz-Jaramillo13}. In some Babcock-Leighton type dynamo models
\citep[e.g.,][]{Chatterjee04,Jiang13}, this correlation exists
because the poloidal field generated by the surface flux transport
can be quickly transported to the tachocline where it gets wound up
by the differential rotation to produce the strong toroidal flux
that emerges in the sunspots of the next cycle.

The strength of the polar fields and the axial dipole moment depend
on the surface flux transport processes -- both the active region
sources (total magnetic flux, polarity separation, and latitude of
emergence) and the surface flows (differential rotation, meridional
flow, and the random convective flows). These processes have been
found to vary systematically with both the phase and the strength of
sunspot cycles.

The transport processes are dominated by the observed surface flows
that include both the large-scale axisymmetric flows (differential
rotation and meridional flow) and the smaller scale non-axisymmetric
flows (granules, supergranules, giant cells, and flows associated
with active regions). These non axisymmetric convective flows are
usually treated as diffusion. Some models also include a decay term
in addition to the observed surface flows. The combined effects of
these transport processes on the emergent sunspot groups impact the
Sun's axial dipole magnetic field in different ways depending on
latitude. While high latitude sunspots typically have more
latitudinal separation between polarities, sunspots emerging closer
to the equator can contribute more to the axial dipole moment by way
of cross-equatorial cancelation.

We note that an important aspect of the magnetic flux transport at
the solar surface is the natural tendency for perturbations in the
sizes of sunspot cycles to produce cycles that continue to grow in
size or  decay in size (with the inability to reverse the polar
fields). On the Sun this tendency must be held in check by some
nonlinear feedback mechanism. We discussed some of the possible
mechanisms -- active region tilt dependent on cycle size, active
region latitude distribution dependent on cycle size, variations in
the meridional flow dependent on cycle size. At this time it is not
clear which, if any, of these mechanisms dominate. It may be that
one mechanism limits the growth while another limits the decay and
the competition between the two keeps sunspot cycles from exhibiting
even more variability.

We now have more that a cycle of reasonably high-resolution and high
temporal cadence observations of the magnetic field and the surface
flows from SOHO/MDI and SDO/HMI. Extending backwards in time we have
over a hundred years of daily records of sunspot group sizes and
locations, as well as knowledge of the Sun¡¯s open magnetic flux
inferred from geomagnetic field measurements (see the review by
Svalgaard, this volume). Looking even further back in time, we have
sunspot number data extending through the Maunder Minimum. Given
this data (and in particular the well-observed transition from large
cycle 22 to small cycle 24), we expect that the evolution of the
Sun's large-scale magnetic field is entering a new stage of
understanding.

\begin{acknowledgements}
We are grateful to the referee for helpful comments on the paper. We
acknowledge the support from ISSI Bern, for our participation in the
workshop on the solar activity cycle: physical causes and
consequences. J.J. acknowledges the financial support by the
National Natural Science Foundations of China (11173033, 11221063,
2011CB811401) and the Knowledge Innovation Program of the CAS
(KJCX2-EW-T07). S.K.S. acknowledges the partial support for this
work by the BK21 plus program through the National Research
Foundation (NRF) funded by the Ministry of Education of Korea. L.G.
acknowledges support from DFG SFB 963 {\it Astrophysical Flow
Instabilites and Turbulence} (Project A18/1) and from EU FP7
Collaborative Project {\it Exploitation of Space Data for Innovative
Helio- and Asteroseismology} (SPACEINN).

\end{acknowledgements}

\clearpage

\begin{figure}[ht!]
\includegraphics[width=\columnwidth]{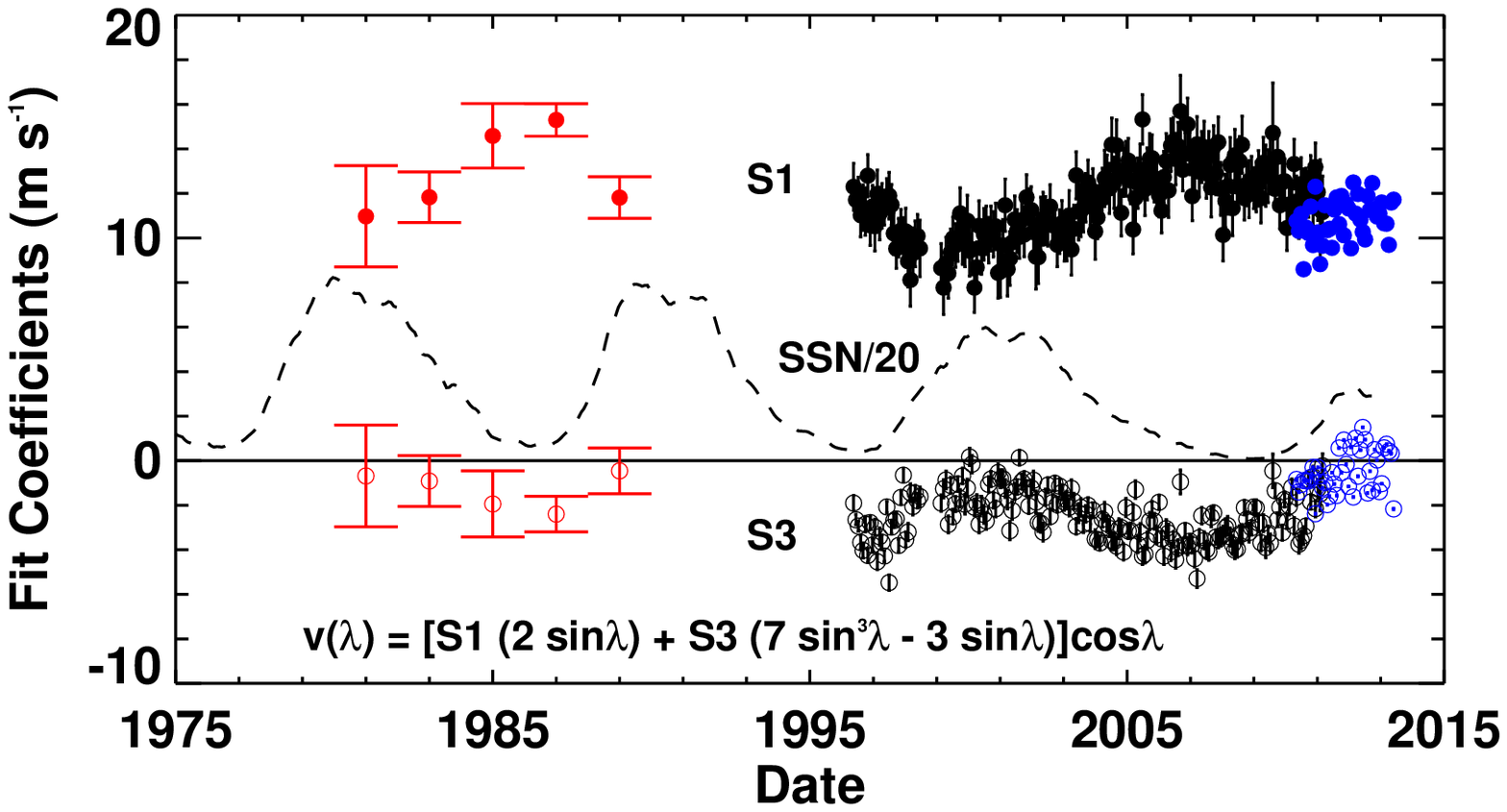}
\caption{The recent history of the polynomial fit coefficients for
meridional flow profiles as measured by the motions of the magnetic
elements. The coefficient S1 is represented by filled circles. S3 is
represented by open circles. The \citet{Komm93} measurements
for 1975-1991 are shown in red. The \citet{Hathaway11}
measurements for individual Carrington rotations (1996-2010) are
shown in black while recent results obtained from SDO/HMI
measurements are shown in blue.} \label{fig:MFprofileHistory}
\end{figure}

\begin{figure*}
\begin{center}
\includegraphics[scale=0.32]{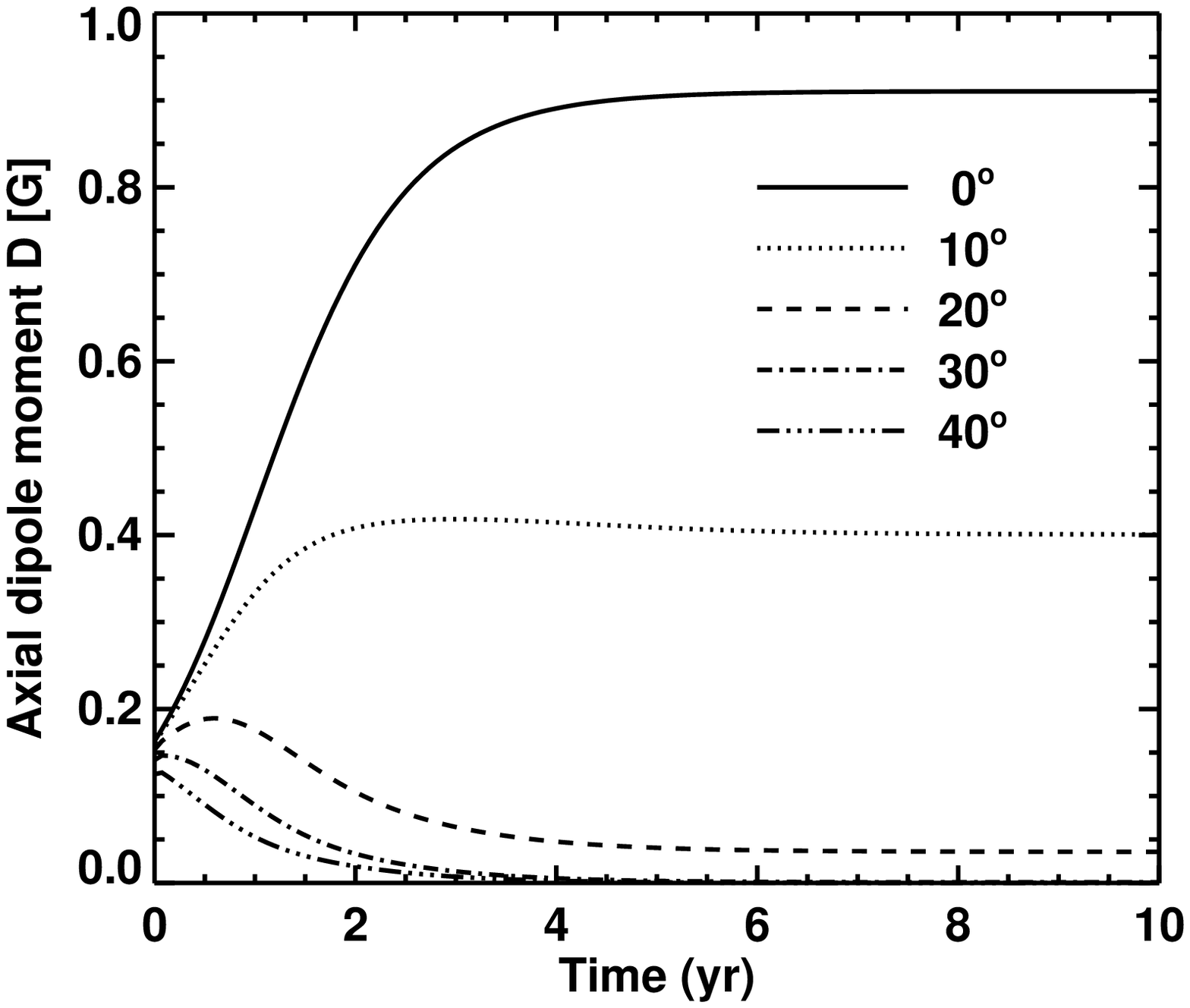}
\includegraphics[scale=0.32]{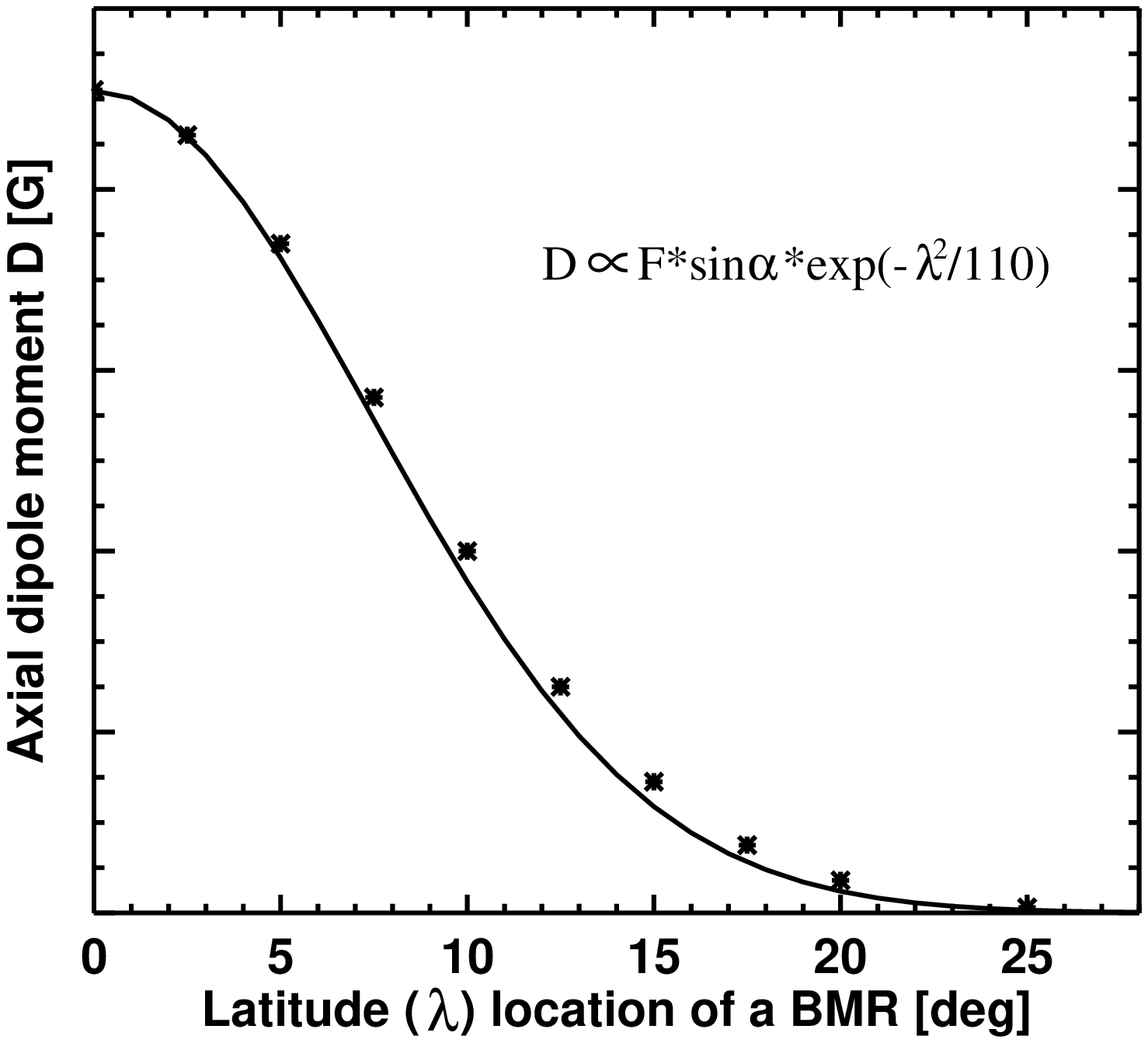}
\end{center}
\caption{The effects of various emergent latitudes for a single BMR
with a total flux of $6\times10^{21}$ Mx and tilt angle 80$^\circ$ (i.e. a nearly N-S oriented dipole)
on the evolution of the Sun's axial dipole moment. Left panel: time
evolution of the axial dipole moment; Right panel: eventual
equilibrium axial dipole field contributed by the single BMR located at
different latitudes.} \label{fig:dip_BMR_lati}
\end{figure*}

\begin{figure*}
\begin{center}
\includegraphics[scale=0.32]{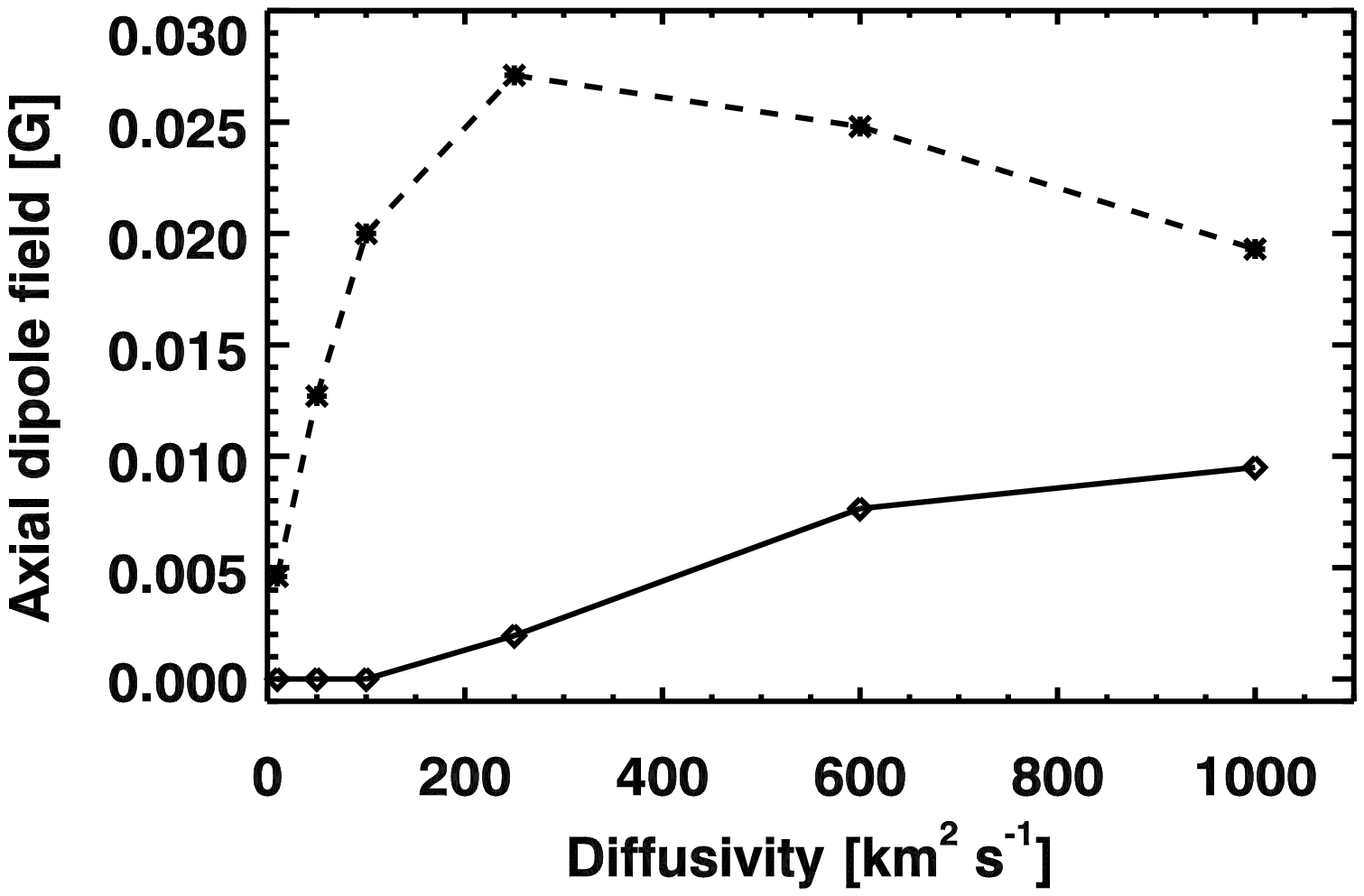}
\includegraphics[scale=0.32]{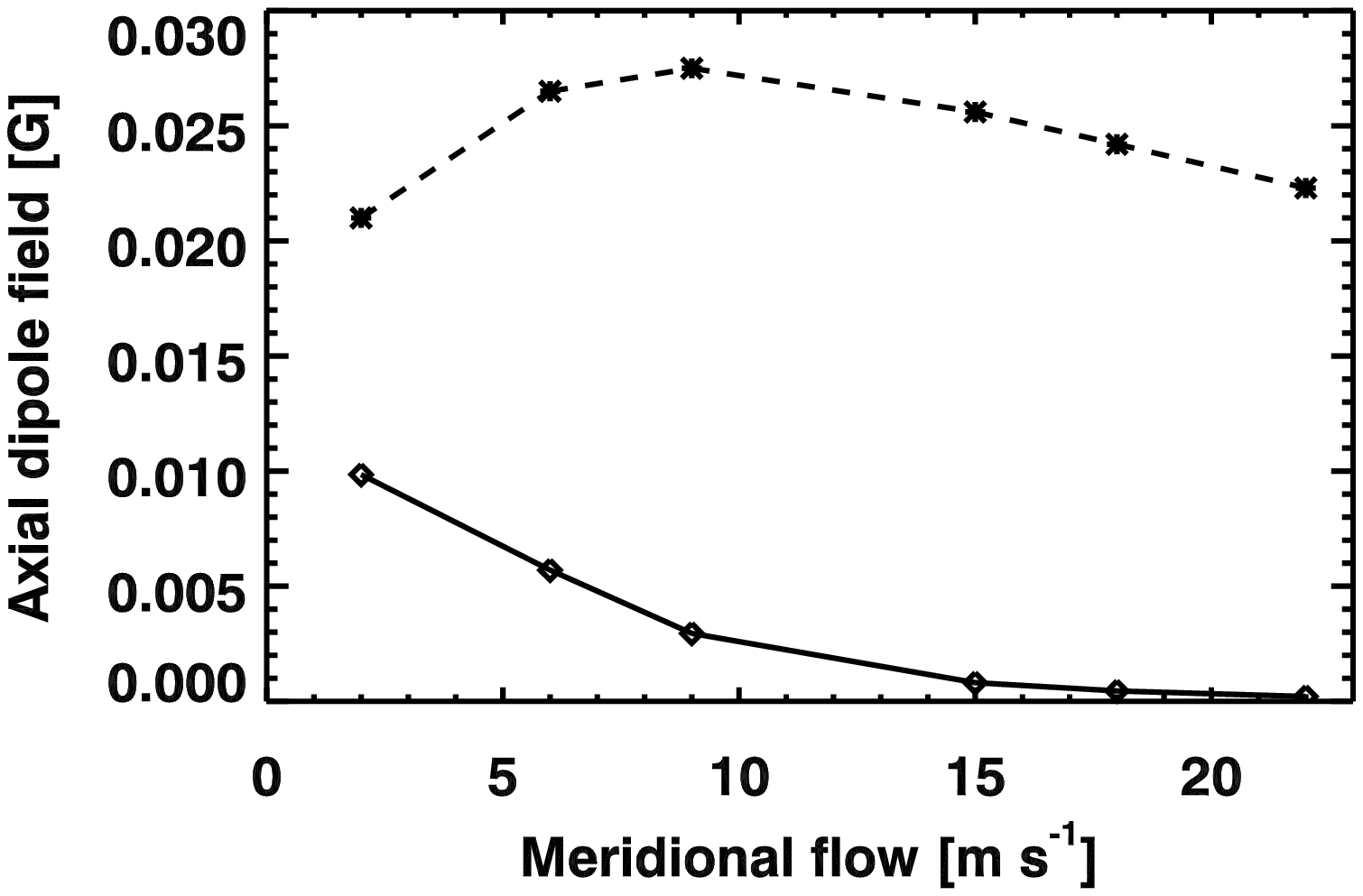}
\end{center}
\caption{Effects of various transport parameters on the eventual
equilibrium axial dipole field for a single BMR with a total flux of
$6\times10^{21}$ Mx and tilt angle 5$^\circ$ deposited at a latitude of 8$^\circ$
(dashed curves) and 18$^\circ$ (solid curves) latitudes. Left panel:
variation of the supergranular diffusivity; Right panel: variation
of the maximum meridional flow.} \label{fig:dip_para}
\end{figure*}

\begin{figure}
\begin{center}
\includegraphics[scale=0.5]{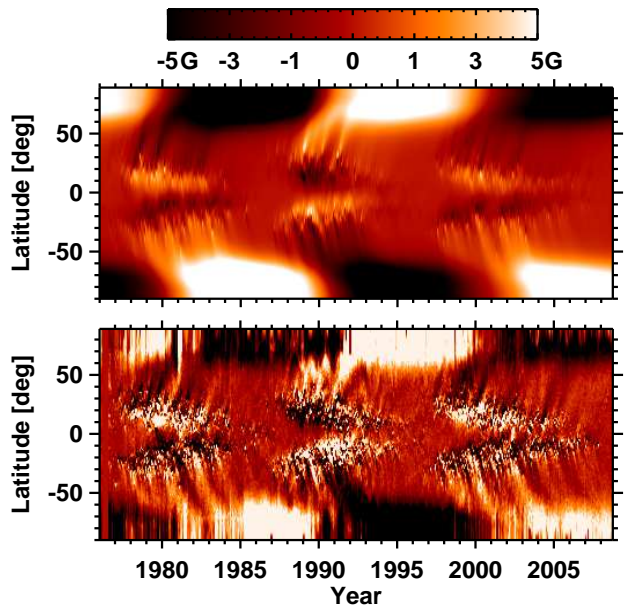}
\end{center}
\caption{Simulated and observed magnetic butterfly diagrams, i.e.,
time-latitude plots of the longitudinally averaged radial magnetic
field at the solar surface. Upper panel: result of the flux
transport simulation based on \citet{Jiang10a}. Lower panel:
evolution of the observed field taken from NSO Kitt Peak synoptic maps.}
\label{fig:MagbutterflyObsSim}
\end{figure}

\begin{figure}
\begin{center}
\includegraphics[scale=0.5]{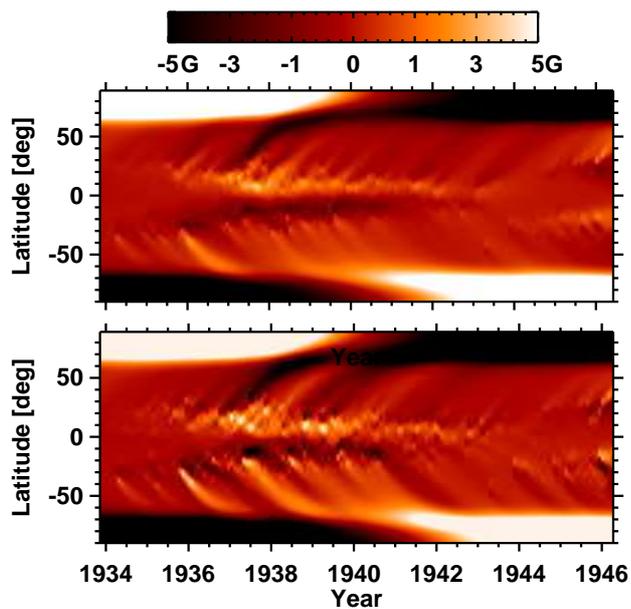}
\end{center}
\caption{Simulated magnetic butterfly diagrams of cycle 17. Upper
panel: case without tilt angle scatter in sunspot groups; Lower
panel: case with tilt angle scatter in sunspot groups, which shows a
more grainy structure in the activity belts and more poleward surges
with both polarities. This figure is based on Figure 3(a) and
4(a) of \citet{Jiang14}.}
\label{fig:MagbutterflyCy17OnOffTiltScatter}
\end{figure}

\begin{figure}
\begin{center}
\includegraphics[scale=0.3]{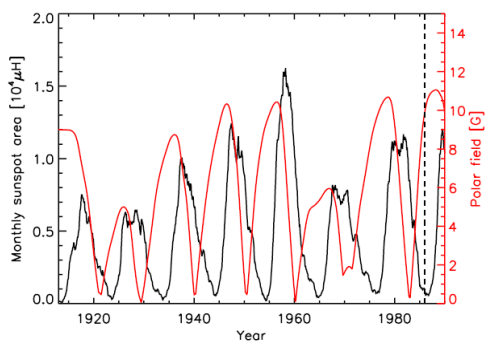}
\end{center}
\caption{Observed sunspot area (black) and average of the unsigned
polar field strength from the flux transport model (red) obtained by
including the nonlinearities in the flux source and with the input
of the RGO sunspot area data during cycles 15 to 21 (from Cameron et
al, 2010).} \label{fig:PolarFieldFluxCJSS10}
\end{figure}

\begin{figure}
\begin{center}
\includegraphics[scale=0.5]{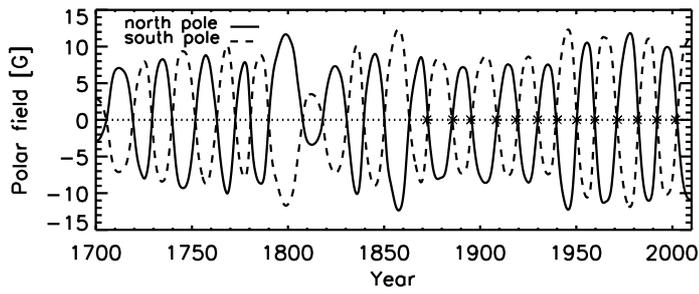}
\end{center}
 \caption{Polar field evolution since 1700
from a flux transport simulation that includes the nonlinearities in
the flux source, with the Wolf sunspot number data used as input
\citep[from][]{Jiang_etal11b}.} \label{fig:PolarFieldJiang11}
\end{figure}

\begin{figure}
\begin{center}
\includegraphics[scale=0.5]{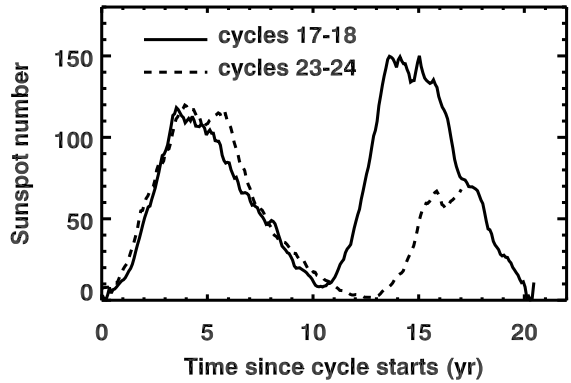}
\end{center}
\caption{Comparison of the time evolution of sunspot number with 12-month running mean
between cycles 17-18 and cycles 23-24. The x-axis denotes the time since the starts of cycles
17 and 23. The two similar cycles 17 and 23 have substantially different subsequent cycles.}
\label{fig:cy17vs23}
\end{figure}

\clearpage
\bibliographystyle{aps-nameyear}      
\bibliography{ISSI_SFT}                
\nocite{*}

\end{document}